\documentclass[12pt]{article}

\usepackage{amsmath}
\usepackage{amssymb}
\usepackage{graphicx}
\usepackage{color}

\usepackage{hyperref}

\usepackage{xr}

\numberwithin{equation}{section}

\newcommand{\gns}{{\rm GNS}}

\newcommand{\A}{D }

\newcommand{\e}{\,{\rm e}}

\newcounter{resultcounter}[section]

\newtheorem{thm}[resultcounter]{Theorem}
\newtheorem{lem}[resultcounter]{Lemma}
\newtheorem{prop}[resultcounter]{Proposition}

\newcommand{\N}{{\mathbb N}}

\renewcommand{\r}{{\rm R}}
\newcommand{\s}{{\rm S}}
\newcommand{\h}{{\cal H}}
\newcommand{\cx}{{\mathbb C}}
\newcommand{\rx}{{\mathbb R}}
\renewcommand{\i}{{\rm i}}

\def\qed{\hfill $\Box$\medskip}

\newcommand{\bbbone}{\mathchoice {\rm 1\mskip-4mu l} {\rm 1\mskip-4mu l}
{\rm 1\mskip-4.5mu l} {\rm 1\mskip-5mu l}}

\renewcommand{\N}{\,\mathcal N}
\newcommand{\Nh}{\widehat N}

\begin{document}

\title{Correlation decay and markovianity \\ 
	in open systems \\
}

\author{Marco Merkli\\
	Department of Mathematics and Statistics\\
	Memorial University of Newfoundland\\
	St. John's, Canada, A1C 5S7\\
 \ \\
\bf Dedicated to the\\
\bf Memory of Gennady P. Berman\\
\bf My Friend and Teacher}
\maketitle
\vspace*{-.3cm}

\medskip
\begin{abstract}
A finite quantum system $\s$ is coupled to a thermal, bosonic reservoir $\r$. Initial $\s\r$ states are possibly correlated, obtained by applying a quantum operation taken from a large class, to the uncoupled equilibrium state. We show that the full system-reservoir dynamics is given by a markovian term plus a correlation term,  plus a remainder small in the coupling constant $\lambda$ uniformly for all times $t\ge 0$. The correlation term decays polynomially in time, at a speed independent of $\lambda$. After this, the markovian term becomes dominant, where the system evolves according to the completely positive, trace-preserving semigroup generated by the Davies generator, while the reservoir stays stationary in equilibrium. This shows that (a) after initial $\s\r$ correlations decay,  the $\s\r$ dynamics enters a regime where both the Born and Markov approximations are valid, and (b) the reduced system dynamics is markovian for all times, even for correlated $\s\r$ initial states.   
\end{abstract}

\section{Introduction}

Open quantum systems are most commonly described by a tensor product Hilbert space ${\mathcal H}_\s\otimes {\mathcal H}_\r$ (system and reservoir) and an interacting Hamiltonian
$$
H_\lambda = H_\s\otimes\bbbone_\r+\bbbone_\s\otimes H_\r+\lambda V
$$ 
where $H_\s$ and $H_\r$ are the system and reservoir Hamiltonians, $\lambda$ is a coupling constant, which we consider to be small in a suitable sense, and $V$ is an interaction operator. In the literature, coupled $\s\r$ systems are sometimes studied when the parts $\s$ and $\r$ are of comparable size. In the current work, we focus on the situation $\dim{\mathcal H}_\s=N<\infty$ and $\dim \mathcal H_\r=\infty$, and a reservoir Hamiltonian $H_\r$ with continuous spectrum. In this sense,  the reservoir $\r$ is much larger than the system $\s$ in our considerations.  The dynamics of an initial state (density matrix) $\rho_{\s\r}$ is given by the Liouville--von Neumann (or Schr\"odinger) equation
\begin{equation*}
\rho_{\s\r}^t = e^{-\i t H_\lambda} \rho_{\s\r} \, e^{\i t H_\lambda}
\end{equation*}
and the reduced system density matrix is obtained by taking the partial trace over the reservoir,
\begin{equation*}
\rho_{\s}^t = {\rm tr}_\r \, \big\{ e^{-\i t H_\lambda} \rho_{\s\r} \, e^{\i t H_\lambda}\big\}.
\end{equation*}
\medskip

{\bf Uncorrelated initial states. } In textbooks and research articles, most often the assumption is made that the initial state is uncorrelated,  that is, of the product form $\rho_{\s\r} = \rho_\s\otimes\omega_\r$. 
Then one can define the {\em dynamical map} acting on system density matrices,
$$
{\mathcal V}_t: \rho_\s\mapsto \rho_\s^t .
$$
$\mathcal V_t$ is the flow or propagator of the system dynamics and of course, it depends on the initial reservoir state $\omega_\r$. For example, if  $\omega_\r=\omega_{\r,\beta}$ is the reservoir equilibrium state at a given temperature $T=1/\beta$, then of course the system dynamics, and hence $\mathcal V_t$, will depend on $T$. Due to the interaction of $\s$ and $\r$, mediated by the operator $V$, the state $\rho_{\s\r}^t$ is generically correlated (not of product form) for $t>0$. This results in the violation of the group property,
$$
\mathcal V_{t+t'}\neq \mathcal V_t  \mathcal V_{t'}.
$$ 
If the reservoir is not much influenced by the coupling to the system  and initially in a stationary state under its own dynamics, then one might intuitively expect that $\rho_{\s\r}^t\approx \rho_\s^t\otimes\omega_\r$. This is called the {\em Born approximation}. If on top, one assumes that the reservoir loses its memory quickly, then one may expect the Markovian property to hold approximately, 
\begin{equation*}
\mathcal V_{t+t'}\approx \mathcal V_t \mathcal V_{t'}\quad \mbox{or}\quad \mathcal V_t \approx e^{t {\mathcal L}},
\end{equation*}
where the generator ${\mathcal L}=\mathcal L(\lambda)$ depends implicitly on the initial reservoir state $\omega_\r$ and $\mathcal L(0) =-\i [H_\s,\cdot\, ]$ generates the uncoupled system dynamics.  It has been known since the seventies (see the founding papers \cite{Davies1, Davies2, VanHove}) that in some generality, the Markovian approximation is valid in the weak coupling, or  van Hove limit. This means that 
\begin{equation}
	\lim_{\lambda\rightarrow 0} \sup_{0\le \lambda^2 t<a} \|\mathcal V_t -e^{t{\mathcal L}_\s(\lambda)}\| =0\qquad \mbox{for any $a>0$},
\label{12-i}
\end{equation}
where the superoperator 
\begin{equation}
	{\mathcal L}_\s(\lambda) = {\mathcal L}_\s +\lambda^2 {\mathcal  K},\qquad {\mathcal  L}_\s = -\i [H_\s, \cdot\, ]
\label{daviesgen}
\end{equation} 
is called the {\em Davies generator}. It generates the CPTP semigroup  $e^{t\mathcal L_\s(\lambda)}$ \cite{CP,DS,MAOP} (completely positive trace preserving). The operator $\mathcal K$, a sum of a Hamiltonian plus a dissipative term, describes the influence of the bath to second order in perturbation \cite{AL,BreuerPetruccione,RivasHuelga}.  From \eqref{12-i} we have an alternative expression for $\mathcal K$, 
\begin{equation}
	\lim_{\lambda\rightarrow 0}\ \mathcal V_{\tau/\lambda^2}\, \circ\, e^{-\tau {\mathcal L}_\s/\lambda^2} = e^{\tau \mathcal K},\qquad \forall \tau\ge 0.
\label{1.2-i}
\end{equation}
The shortcoming of \eqref{12-i} is that it guarantees the accuracy of the Markovian approximation only for bounded values of $\lambda^2 t$. Said differently, \eqref{12-i} shows that  the Markovian approximation is guaranteed to hold for $t\rightarrow\infty$ only if at the same time, one takes $\lambda\rightarrow 0$ in a way such that $\lambda^2 t$ stays finite. This deficiency of \eqref{12-i} was overcome in \cite{KM2,MAOP,Markov1, Markov2},  where it is proven, under suitable assumptions, that
\begin{equation}
	\sup_{t\ge 0}\big \| \mathcal V_t -e^{t({\mathcal L}_\s+\lambda^2\mathcal K)}\big\| \le C |\lambda|^{1/4}, 
	\label{14-i}
\end{equation}
provided $|\lambda|\le \lambda_0$ for some $\lambda_0>0$. The Markovian approximation is guaranteed to be accurate to $O(|\lambda|^{1/4})$ {\em uniformly in time} $t\ge 0$, for all small, fixed $\lambda$. The error bound $|\lambda|^{1/4}$ is derived for the least amount of regularity of the interaction operator necessary for the proof -- under a more stringent (analyticity) assumption, it can be improved to $|\lambda|^2$ (see \cite{KM2}). In the recent work \cite{Burgarthetal} it is shown that the validity of the Markovian approximation for arbitrarily large {\em finite} times does not imply the validity for all times. There the authors consider the zero temperature, dissipative Janes-Cummings model in the singular coupling limit.  Given any time $t_0>0$, they explicitly design SR interactions (form factors $g(k)$) in such a way that the system (qubit) dynamics is {\em exactly Markovian} up to $t_0$, but for $t>t_0$ the dynamics becomes non-Markovian. Even though this example uses the singular coupling limit ($g(k)$ not square integrable), the result indicates that the passage from \eqref{12-i} to \eqref{14-i} is not a mere technicality. In an approach different from the above, a polymer expansion for initially factorized system-reservoir states was established in \cite{DK}.

\medskip

{\bf Correlated initial states.} The assumption of initially uncorrelated SR states may not be physically realistic. It requires to bring together, at time $t=0$, a system and a reservoir which at earlier times did not interact (or which by some measurement or dynamical fluke would happen to be in a product state). One should then ask about the dynamics of correlated initial states, that is, ones which are not of product form, see \cite{Fleming, Modi, R2,Tasakietal, Yuasaetal} and the references in there. For a given correlated initial state $\rho_{\s\r}$, the reduced system dynamics  $\rho_\s^t$ is still well defined as above. However, one cannot define the dynamical map $\mathcal V_t$ any longer, as different $\rho_{\s\r}$ can have the same marginal $\rho_\s$ (say, due to different $\s\r$ correlations). Those different initial $\s\r$ states evolve to states at time $t>0$ which have different system marginals, which means that the initial marginal cannot be mapped consistently to the marginal at a later time.  In our investigation, we {\em start} with a correlated initial state $\rho_{\s\r}$ and describe its well defined reduced system dynamics $t\mapsto \rho^t_\s$. The opposite approach is also studied in the literature, starting with an initial system state $\rho_\s$, then mapping it to a $\s\r$ state  by an {\em assignment map} $\rho_\s\mapsto \rho_{\s\r}$. The initial correlation is then encoded in the assignment map, whose structure can be analyzed  \cite{Alicki, Dominyetal, DominyLidar, Pechukas, VacchiniAmato}.

\medskip

{\bf Outline of main results. } We consider a class of open systems for which the Markov approximation \eqref{14-i} for uncorrelated initial states, $\rho_\s\otimes\omega_{\r,\beta}$, was proven in \cite{Markov2}. Here, $\omega_{\r,\beta}$ is the thermal equilibrium state of the reservoir at temperature $T=1/\beta>0$.\footnote{This is not a necessary feature for our approach to work, one may also take non-equilibrium stationary states, as we will explain elsewhere.} Our main result, Theorem \ref{theorem1}, gives an expansion of the full $\s\r$ dynamics with correlated initial states $\rho_{\s\r}$. It can be expressed as follows -- see after Theorem \ref{theorem1} for more in-depth and precise statements. We show that 
\begin{equation}
	\rho^t_{\s\r}=\big(e^{t{\mathcal L}_\s(\lambda)}{\rm tr}_\r\{\rho_{\s\r}\}\big)\otimes\omega_{\r,\beta} + \chi(t,\lambda) + O(|\lambda|^{1/4}),
	\label{r2i}
\end{equation}
where $\mathcal L_\s(\lambda)$ is the Davies generator \eqref{daviesgen} and $\chi(t,\lambda)$ describes the dynamics of the $\s\r$ correlations. The relation\eqref{r2i} is valid for all $|\lambda|\le \lambda_0$, some $\lambda_0>0$, with a remainder independent of time $t\ge0$. It is to be interpreted in a weak sense, that is, when applied to operators from a $\s\r$ observable algebra $\mathcal O$, \eqref{1.14}. The initial states are taken from a class of states obtained by applying (very general) quantum operations on the uncoupled $\s\r$ equilibrium state $\rho_{\s,\beta}\otimes\omega_{\r,\beta}$ (see \eqref{c62}). We show the following results:
\begin{itemize}
\item[--] For $\rho_{\s\r}=\rho_\s\otimes\omega_{\r,\beta}$ (with $\rho_\s$ any system density matrix), we have $\chi(t,\lambda)=0$. This shows that {\em the Born approximation holds} and recovers the Markovianity result \eqref{14-i}.  

\item[--] $\chi(t,\lambda)$ also vanishes when applied to observables $A_\s\otimes\bbbone_\r$ of $\s$ alone, for any initial $\rho_{\s\r}$. This means that the initial correlations only influence the dynamics of observables which involve the reservoir. In particular, this shows that {\em the Markovian approximation is also valid for correlated initial states}, and for all times $t\ge0$. 

\item[--] In the presence of initial correlations, the Born approximation ($\rho_{\s\r}^t\approx \rho_\s^t\otimes\omega_{\r,\beta}$) is obviously not correct for small times. However, we show that $\chi(t,\lambda)\sim 1/(1+t^3)$, which implies that initial correlations decay, and after that, the $\s\r$ state enters a regime where the Born approximation is valid, even for correlated initial states. This is a statement about the {\em full} $\s\r$ state dynamics. In contrast, as explained above, the {\em reduced} system state is approximated by the Markovian semigroup for all, even small, times.
\end{itemize}

\section{Model and main results}

\subsection{Model}

We consider an $N$-level system interacting with a reservoir modeled by a continuous family (field) of quantum harmonic oscillators, and we will largely follow the notation of \cite{Markov1}. The system and reservoir Hamiltonians are given by
\begin{equation}
	H_\s = \sum_{j=1}^N E_j |\phi_j\rangle\langle\phi_j| \mbox{\quad and\quad}	H_\r = \int_{{\mathbb R}^3}\omega(k) a^*(k)a(k)d^3k 
	\label{n1}
\end{equation}
respectively, acting on the Hilbert space
\begin{equation}
{\mathcal H}_{\s\r} = {\mathbb C}^N \otimes	{\mathcal F}.
\label{2.2}
\end{equation}
We often simply write $H_\s$ for $H_\s\otimes\bbbone$ and so on. The $E_j$  are the system energies and  $\omega(k)\ge 0$ is the frequency of the mode $k$ ($\hbar=1$), which we take for definiteness to follow the dispersion {\em e.g.} of photons,
\begin{equation}
	\omega(k)= |k|.
\end{equation}
The energy eigenstates $\phi_j$ form an orthonormal basis of the system Hilbert space ${\mathbb C}^N$ and the (momentum representation) creation and annihilation operators satisfy the canonical commutation relations $[a(k),a(\ell)]=0=[a^*(k), a^*(\ell)]$ and $[a(k),a^*(\ell)]=\delta(k-\ell)$ (Dirac delta $k,\ell\in{\mathbb R}^3$). The reservoir Hilbert space is the (Bosonic) Fock space
\begin{equation}
{\mathcal F} = \bigoplus_{n\ge 0} \, L_{\rm sym}^2({\mathbb R}^{3n}, d^{3n}k), 
\label{F}
\end{equation}
built over the single-particle space $L^2({\mathbb R}^{3},d^3k)$,  the square integrable scalar valued functions.  The full, interacting Hamiltonian is given by 
\begin{equation}
	\label{n2}
H_\lambda = H_\s +H_\r +\lambda G\otimes\varphi(g),
\end{equation}
where $\lambda\in\mathbb R$ is the coupling constant, $G^*=G$ is any hermitian system matrix and 
\begin{equation}
\varphi(g) =\frac{1}{\sqrt 2}\big[a^*(g)+a(g)\big],\quad a^*(g)=\int_{{\mathbb R}^3} g(k) a^*(k)d^3k, \quad a(g) = [a^*(g)]^*
\label{c60}
\end{equation}
are the field, creation and annihilation operators, respectively. 
The function $g(k)\in L^2({\mathbb R}^3, d^3k)$ in the interaction \eqref{n2} is called the form factor -- generally, when we consider $a^*(f)$, $a(f)$ we also call $f$ test functions. 

The above is a paradigmatic model for open quantum systems. For $N=2$ it is called the spin-Boson model. As presented here, it is the continuous mode (or thermodynamic, infinite-volume) limit of a system described by a discrete set of oscillators with Hamiltonian 
\begin{equation}
H_\r' = \sum_{k} \omega_k a^*_ka_k
\label{H'}
\end{equation} 
and field operator in the interaction given by $\varphi'(g) =\sum_k g_k a^*_k+{\rm h.c.}$ In the theoretical physics literature, the limit of continuous values of $k$ is often performed at the end of calculations of various expressions,  such as transition probabilities.  In this limit of a continuum of reservoir modes one can analyze irreversible dynamical effects (like time decay).  In the present work, we start off directly with the reservoir having a continuum of frequencies, \eqref{n1}. 
\medskip

In contrast to the common assumption of initially uncorrelated system-reservoir states of the factorized form $\rho_\s\otimes\rho_\r$, we are concerned here with initial states obtained by correlating the system with the reservoir.  The system Gibbs equilibrium density matrix  at temperature $T=1/\beta$, 
\begin{equation}
\rho_{\s,\beta} = Z_{\s,\beta}^{-1}\,  e^{-\beta H_\s}, 
\label{c61}
\end{equation}
$Z_{\s,\beta}={\rm tr}_\s e^{-\beta H_\s}$, is a well defined density matrix. One can represent $\rho_{\s,\beta}$ as a vector state $|\Omega_\s\rangle\langle \Omega_\s|$ in the enlarged Hilbert space ${\mathbb C}^N\otimes{\mathbb C}^N$ as follows. For any $A_\s\in{\mathcal B}({\mathbb C}^N)$ (bounded operators), 
\begin{equation}
{\rm tr}_{{\mathbb C}^N} \big( \rho_{\s,\beta} A_\s\big) = \langle \Omega_\s, \pi_\s(A_\s)\Omega_\s\rangle_{{\mathbb C}^N\otimes{\mathbb C}^N},
\end{equation}
where $\pi(A_\s)=A_\s\otimes\bbbone_{{\mathbb C}^N}$ and $\Omega_\s\in {\mathbb C}^N\otimes{\mathbb C}^N$ is a normalized vector (explicitly given in  \eqref{n47.1}). This purification representation is sometimes very helpful. For the continuous mode reservoir the situation is a bit more cumbersome. Its equilibrium state cannot be expressed as a density matrix acting on $\mathcal F$, because $e^{-\beta H_\r}$ is not trace class ($H_\r$ has continuous spectrum). Rather, the reservoir equilibrium state is constructed as the limit of equilibrium states $\propto e^{-\beta H'_\r}$, where $H'_\r$, \eqref{H'} is trace class. The limit state is expressed as a positive, linear functional $\omega_{\r,\beta}$ on reservoir operators as follows. Let 
\begin{equation}
	L^2_0 = L^2({\mathbb R}^3,d^3k)\cap L^2({\mathbb R}^3,|k|^{-1}d^3k).
\end{equation}
For $f,g\in L^2_0$ we have \cite{BR,MLnotes}
\begin{equation}
\omega_{\r,\beta}(a(f))=\omega_{\rm \r,\beta}(a^*(f))=0,\quad 	\omega_{\r,\beta}\big( a^*(f) a(g)\big) = \langle g,(e^{\beta |k|}-1)^{-1} f\rangle,
	\label{2pt}
\end{equation} 
where  $\langle f,g\rangle = \int_{{\mathbb R}^3} \bar f(k) g(k) d^3k$. Expectations of arbitrary polynomials in creation and annihilation operators are calculated using Wick's theorem \cite{BR} and the expectation of a unitary Weyl operator
\begin{equation}
	W(f) = e^{\i \varphi(f)}, \quad f\in L^2_0,
	\label{1.8}
\end{equation}
is 
\begin{equation}
	\omega_{\r,\beta}\big(W(f)\big) = e^{ -\tfrac14 \langle f,\coth(\beta|k|/2)f\rangle}.
	\label{1.10}
\end{equation}
It is very convenient to use a purification of $\omega_{\r,\beta}$, that is, to represent this state as a {\em vector state} in a new Hilbert space ${\mathcal H}_\r$ (different from ${\mathcal F}$, \eqref{F}), in which the operators $a^\sharp(f)$ (where $a^\sharp=a$ or $a^\sharp=a^*$) and $W(f)$ are represented by operators 
\begin{equation}
a^\sharp(f) \mapsto \pi_\r\big(a^\sharp(f)\big),\qquad W(f)\mapsto \pi_\r(W(f)).
\label{3.12}
\end{equation}
Here, $\pi_\r$ is a linear map, sending operators on $\mathcal F$ to operators on ${\mathcal H}_\r$ and satisfying $\pi_\r(X^*)=\pi_\r(X)^*$ and $\pi_\r(XY)=\pi_\r(X)\pi_\r(Y)$, {\em i.e.}, $\pi_\r$ is a $*$ algebra representation. Let $A_\r$ be an arbitrary finite sum of products of creation and annihilation operators and Weyl operators acting on $\mathcal F$. Then we have 
\begin{equation}
\omega_{\r,\beta}(A_\r) = \langle\Omega_\r, \pi_\r(A_\r)\Omega_\r\rangle_{{\mathcal H}_\r}
\label{2.15}
\end{equation}
for a normalized vector $\Omega_\r\in{\mathcal H}_\r$. In this sense, $\omega_{\r,\beta}$ is represented by the vector $\Omega_\r$. The explicit form of ${\mathcal H}_\r$, $\pi_\r$ and $\Omega_\r$ is well known (Araki-Woods representation \cite{AW}), we give them in Section \ref{purifsect}. 

The purification of the uncoupled joint equilibrium state $\rho_{\s,\beta}\otimes\omega_{\r,\beta}$ is obtained by taking the tensor product: For operators $A$ which are are arbitrary finite sums of products of bounded operators on ${\mathbb C}^N$,  creation and annihilation operators and Weyl operators (with test functions in $L^2_0$), we have
\begin{equation}
\rho_{\s,\beta}\otimes\omega_{\r,\beta}(A) = \langle \Omega_{\s\r,\beta,0}, \pi(A) \Omega_{\s\r,\beta,0}\rangle_{{\mathcal H}_{\gns}},
\end{equation}
where (GNS stands for {\em Gelfand-Naimark-Segal})
\begin{equation}
{\mathcal H}_{\gns} = {\mathbb C}^N\otimes{\mathbb C}^N\otimes {\mathcal H}_\r,\quad   \pi= \pi_\s\otimes\pi_\r\quad \mbox{and}\quad \Omega_{\s\r,\beta,0} = \Omega_\s\otimes\Omega_\r.
\label{3.17}
\end{equation}

{\bf Remarks about the use of unbounded operators. } In the mathematical literature on $C^*$  and $W^*$ dynamical systems one considers observables to be elements of the Weyl algebra, which are {\em bounded} operators. However, in the physics literature, it is more common to work with creation and annihilation operators $a^*(f)$ and $a(f)$, as they carry direct physical meaning. In the mathematical $W^*$ algebraic setting, the equilibrium state of the infinitely extended reservoir state is given as the expectation functional \eqref{1.10} defined on the Weyl algebra. Alternately, it is expressed by \eqref{2.15} for $A_\r=W(f)$. The representation $\pi_\r$ is regular, which means that $-i\partial_\alpha|_{\alpha =0}\pi_\r(W(\alpha f))$ is a well defined self-adjoint operator on $\h_\r$, interpreted as the represented field operator ({\em c.f.}~\eqref{c60}).  It is then natural to extend the domain of $\pi_\r$ by defining $\pi_\r(\varphi(f))=-i\partial_\alpha|_{\alpha =0}\pi_\r(W(\alpha f))$. Similarly one defines $\pi_\r(a^\sharp(f))$ and generally the action of $\pi_\r$ on any polynomial of creation and annihilation operators. (Here, $a^\sharp$ stands for $a$ or $a^*$.) It then makes perfect sense to define $\omega_{\r,\beta}$ on such polynomials by
$$
\omega_{\r,\beta}(a^{\sharp}(f_1)\cdots a^{\sharp}(f_n)) = \langle\Omega_\r, \pi_\r( a^{\sharp}(f_1))\cdots \pi_\r(a^{\sharp}(f_n))\Omega_\r\rangle_{\h_\r}. 
$$
Of course, one must verify that the vector $\Omega_\r$ is in the domain of the unbounded operator $\pi_\r( a^{\sharp}(f_1))\cdots \pi_\r(a^{\sharp}(f_n))$. This is done using the explicit form of $\Omega_\r$ and the operators involved; see  \eqref{3.9}, \eqref{c45} and also Section \ref{sec2.4}. In the present work, we consider reservoir observables which {\em are} Weyl operators (for reasons explained after Proposition \ref{prop1.1}), however, we allow for $\s\r$ correlation operators to be made from polynomials in creation and annihilation operators or even exponentials thereof, written as
\begin{equation}
\label{XX}
\exp \big[ i \sum_{r=1}^R B_r\otimes a^\sharp(f_r)\big] = \sum_{n\ge 0} \frac{i^n}{n!} \big[ \sum_{r=1}^R B_r\otimes a^\sharp(f_r)\big]^n
\end{equation}
where $R\in{\mathbb N}$, $B_r\in{\mathcal B}({\mathbb C}^N)$ and $f_r\in L^2( \mathbb R^3)$ are suitable test functions. The operator on the left side of \eqref{XX} is defined by the series on the right side, which converges strongly on a dense set of vectors in $\h_{\s\r}$, \eqref{2.2} (see Lemma \ref{lemma1.0}). One then also defines 
\begin{equation}
\label{XXX}
\pi\big( \exp \big[ \sum_{r=1}^R B_r\otimes a^\sharp(f_r)\big]\big) = \sum_{n\ge 0} \frac{1}{n!} \big[ \sum_{r=1}^R \pi_\s( B_r)\otimes \pi_\r( a^\sharp(f_r))\big]^n,
\end{equation}
where right hand side is strongly convergent on dense set of vectors in $\h_{\rm GNS}$, see Lemma \ref{lemma4.4}.

\subsection{Observable algebra $\mathcal O$ and correlation algebra $\mathfrak C$}
\label{sec:oa}

We introduce two spaces of single reservoir particle states (`test functions').

\begin{itemize}
\item[(a)] The space of observable test functions is $L^2_{\rm obs}\subset L^2({\mathbb R}^3, d^3k)$, consisting of all  functions which are three times continuously differentiable in the radial variable $|k|>0$ and such that $|f(k)|\le C |k|^{-q}$ for  some $q>7/2$ provided $|k|$ is large enough. Moreover, the infra-red behaviour of $f$ is as follows. There is a $\kappa_0>0$ such that for $|k|< \kappa_0$, we have  $g(k)=|k|^p h(k)$ where either $p>2$ and $h$ is a three times continuously differentiable function, or $p=-\tfrac12, \tfrac12,\tfrac32$ and $h(k)=h_0\in\cx$ is constant or $h(k)=h(|k|^2)\in\rx$ is  real,  radial and three times continuously differentiable.

\item[(b)] The space of correlation operator test functions is $L^2_{\rm cor}$, consisting of all  functions $f\in L^2_{\rm obs}$ such that $e^{\beta|k|} f\in L^2({\mathbb R}^3,d^3k)$, where $\beta$ is the inverse temperature. 
\end{itemize}

 The {\bf observable algebra} is defined as
\begin{equation}
{\mathcal O} = {\mathcal B}({\mathbb C}^N)\otimes{\mathfrak W}_0(L^2_{\rm obs})
\label{1.14}
\end{equation}
where ${\mathcal B}({\mathbb C}^N)$ is the algebra of linear operators on ${\mathbb C}^N$ and 
where ${\mathfrak W}_0(L^2_{\rm obs})$ is the algebra consisting of all finite sums and products of Weyl operators $W(f)$ with $f\in L^2_{\rm obs}$.\footnote{It is not hard to extend our results to the case when $\mathcal O$ is the norm closure of \eqref{1.14} ({\em i.e.}, for  ${\mathfrak W}_0(L^2_{\rm obs})$ replaced by the $C^*$-algebra ${\mathfrak W}(L^2_{\rm obs})$), see the discussion after Theorem \ref{theorem1}.} Next, introduce the polynomial algebra
\begin{equation}
{\mathcal P} = {\rm Span}\big\{B\otimes a^{\sharp}(f_1)\cdots a^{\sharp}(f_n)\ :\ B\in {\mathcal B}({\mathbb C}^N),\ n\in{\mathbb N}, \ f_j\in L^2_{\rm cor}	\big\},
\label{c37}
\end{equation}
where ${\rm Span}$ is the linear span (finite complex linear combinations) and $a^{\sharp}$ denotes either of $a$ or $a^*$, individually in each factor in the product in \eqref{c37}. The set $\mathcal P$ consists of all polynomials of creation and annihilation operators with coefficients in ${\mathcal B}({\mathbb C}^N)$. Furthermore, define the set of operators 
\begin{equation}
{\mathcal X} = {\rm Span}\big\{  e^{\i \sum_{r=1}^R B_r\otimes a^\sharp(f_r) }\ :\ R\in{\mathbb N},\ B_r\in{\mathcal B}({\mathbb C}^N),\ f_r\in L^2_{\rm cor}\big\}. 
\label{c36}
\end{equation}
The {\bf correlation algebra} is defined as
\begin{equation}
{\mathfrak C} ={\rm Span}\big\{ K_1\cdots K_n\ :\  n\in{\mathbb N}, \ K_j\in{\mathcal X} \cup{\mathcal P}\big\}.
\label{c35-1}
\end{equation}
While the observable algebra ${\mathcal O}$ consists of {\em bounded} operators on ${\mathcal H}_{\s\r}$,  the elements $K_j$ in \eqref{c35-1} are  generally {\em unbounded} operators. One must be careful about the definition of the product of such operators, showing up in \eqref{c35-1}. See Proposition \ref{prop1.1}.

\subsection{Initial states}

\medskip

We use the notation $\rho(A) = {\rm tr}(\rho A)$ for a density matrix $\rho$ and an observable $A$. Let $K_\alpha\in\mathfrak C$ be Kraus operators satisfying
\begin{equation}
\label{kraus1}
\rho_{\s,\beta}\otimes\omega_{\r,\beta}\Big( \sum_{\alpha=1}^\nu K^*_\alpha K_\alpha\Big)=1,
\end{equation}
where  $\nu\in{\mathbb N}$. We take initial states of the form
\begin{equation}
\rho_{\s\r}(A) =
 \sum_{\alpha=1}^{\nu} \rho_{\s,\beta}\otimes \omega_{\r,\beta}\big( K^*_\alpha A K_\alpha\big), \qquad A\in\mathcal O.
\label{c62}
\end{equation}
The summand of \eqref{c62} is a short hand notation  
for $\langle\Omega_{\s\r,\beta,0}, \pi(K^*_\alpha)\pi(A)\pi(K_\alpha)\Omega_{\s\r,\beta,0}\rangle_{\h_{\rm GNS}}$. One may extend the results to $\nu=\infty$, see the remarks after Theorem \ref{theorem1}.  The condition \eqref{kraus1} guarantees that $\rho_{\s\r}$ is properly normalized, $\rho_{\s\r}(\bbbone)=1$. The initial reduced system state is defined by the relation
\begin{equation}
\rho_\s(A_\s) = \rho_{\s\r}\big(A_\s\otimes\bbbone_\r\big),\quad \forall A_\s\in{\mathcal B}({\mathbb C}^N). 
\label{redin}
\end{equation}	
The full $\s\r$ state at time $t$ is given by 
\begin{equation}
\rho_{\s\r} ( e^{\i t H_\lambda} A e^{-\i t H_\lambda}) = \sum_{\alpha=1}^\nu \rho_{\s,\beta}\otimes\omega_{\r,\beta}\big(K^*_\alpha\, e^{\i t H_\lambda} A e^{-\i t H_\lambda}K_\alpha\big),\qquad A\in\mathcal O.
\label{3.21}
\end{equation}
Technically, it is defined in a standard way as follows (see Section \ref{s2.2}). Use the Dyson series to define the dynamics $e^{\i t H_\lambda} \cdot e^{-\i t H_\lambda}$ starting from the uncoupled dynamics $e^{\i t H_0} \cdot e^{-\i t H_0}$. Then apply the representation map $\pi$ \eqref{3.17} to the Dyson series. One shows that the resulting series converges, and moreover, that the limit equals $e^{\i t L_\lambda}\pi(\cdot)e^{-\i t L_\lambda}$, where $L_\lambda$ is a self-adjoint operator on ${\mathcal H}_{\gns}$, called the {\em Liouville operator}. The summand of \eqref{3.21} is defined as
\begin{equation}
\rho_{\s,\beta}\otimes\omega_{\r,\beta}\big(K^*_\alpha\, e^{\i t H_\lambda} \cdot e^{-\i t H_\lambda}K_\alpha\big) = \langle \Omega_{\s\r,\beta,0}, \pi(K_\alpha^*) e^{\i t L_\lambda} \pi(\cdot) e^{-\i t L_\lambda}\pi(K_\alpha)\Omega_{\s\r,\beta,0}\rangle_{{\mathcal H}_{\gns}}.
\label{3.22}
\end{equation}
We prove the following result in  Section \ref{sec2.4}: 
\begin{prop}
	\label{prop1.1}
	The functional $\rho_{\s\r}(e^{\i t H_\lambda}\cdot e^{-\i t H_\lambda})$ given by \eqref{3.21} is a well defined state (positive, linear, normalized functional) on $\mathcal O$, for any choice of the $K_j\in{\mathfrak C}$.
\end{prop}

One can show that $\rho_{\s\r}$ is also defined (finite) on observables $O$ which are products of polynomials $B\otimes a^{\sharp}(f_1)\cdots a^{\sharp}(f_n)$ and operators $e^{\i \sum_{r=1}^R B_r\otimes a^\sharp(f_r)}$ with test functions $f_j\in L^2_{\rm obs}$. This is a larger class than ${\mathcal O}$. However, it will be more difficult to give a proof of our main result, Theorem \ref{theorem1}, for unbounded observables. In the proof, we use the time uniform estimate $\|e^{\i t H_\lambda} A e^{-\i t H_\lambda}\|=\|A\|$, valid for {\em bounded} observables $A$. For unbounded $A$ (say a product of creation and annihilation operators) one has to find time uniform bounds of $e^{\i t H_\lambda} A e^{-\i t H_\lambda}$ in a weak sense (on suitable functionals). This requires a more complicated analysis which we do not address it here. 
\medskip

The collection of initial states does not depend the state $\rho_{\s,\beta}$ in the definition \eqref{c62}. Indeed, suppose we had put an arbitrary system density matrix $\sigma_\s$ in the place of $\rho_{\s,\beta}$ in \eqref{c62}. Then, since  $\sigma_\s = K\rho_{\s,\beta} K^*$ for $K=\sqrt{\sigma_\s}\,   (\rho_{\s,\beta})^{-1/2}$, we obtain the same class of initial states because changing the system reference state simply amounts to a change in the Kraus operators. The same is not true for the role of $\omega_{\r,\beta}$ in \eqref{c62}. If we replace there $\omega_{\r,\beta}$ by $\omega_{\r,\beta}(K^*\cdot K)$ for some reservoir operator $K\in{\mathfrak C}$ (bounded or unbounded), then we still obtain the same class of initial states. The set of such states $\omega_{\r,\beta}(K^*\cdot K)$ are called {\em normal states} w.r.t. $\omega_{\r,\beta}$, they form the {\em folium} of $\omega_{\r,\beta}$.  However, not all states of the reservoir are of this form; for instance, $\omega_{\r,\beta'}$ is not normal w.r.t. $\omega_{\r,\beta}$ unless $\beta=\beta'$.  Normal states differ from each other only `quasilocally'. We refer to \cite{BR,Haag} for a more precise discussion of this point.
\medskip

{\em Example. } An admissible initial state is for instance 
$$
\rho_{\s\r}=\Lambda(\rho_{\s,\beta}\otimes\omega_{\r,\beta}) = \frac{1}{Z} \, e^{{\mathcal E}}\big( \rho_{\s,\beta}\otimes\omega_{\r,\beta}\big)\, e^{{\mathcal E}^*},
$$
where $Z$ is a normalization factor and  ${\mathcal E}=\sum_{j} B_j\otimes a^*(f_j) +\sum_k D_k\otimes a(f_k)$ is an arbitrary expression linear in the creation and annihilation operators. This is a state in which $\s$ and $\r$ are entangled, {\em c.f.} \cite{Yuasaetal}.

\subsection{The main result}

Assumptions.
\begin{itemize}
	\item[(A1)] {\em Smoothness of the form factor.}
We assume  that the form factor $g(k)$ in the interaction \eqref{n2} is four times continuously differentiable in the raidal variable $|k|>0$ and that $|g(k)|\le C |k|^{-q}$ for some $q>3/2$, for large enough $|k|$. Moreover, the infra-red behaviour of $g$ is characterized as follows: There is a $\kappa_0>0$ such that for $|k|\le \kappa_0$, we have  $g(k)=|k|^p h(k)$ where either $p>3$ and $h$ is a four times continuously differentiable function, or $p=-\tfrac12, \tfrac12, \tfrac32, \tfrac 52$ and $h(k)=h_0\in\cx$ is constant or $h(k)=h(|k|^2)\in\rx$ is  real, radial and four times continuously differentiable.

\item[(A2)] 
\begin{itemize}
	\item[(a)] {\em Fermi Golden Rule Condition}: The reservoir spectral density is defined as
\begin{equation}
	J(\omega) = \tfrac{1}{2}\pi \omega^2\int_{S^2} |g(\omega,\Sigma)|^2 d\Sigma, \qquad \omega\ge 0,
	\label{specdens}
\end{equation}
where $g(k)$ is the form factor in the interaction \eqref{n2} and the integral is over the (polar and azimuthal) angles. We assume that 
\begin{equation}
	\langle \phi_m, G\phi_n\rangle J(|E_m-E_n|) \neq 0 \qquad \mbox{for all system energies $E_m\neq E_n$},
	\label{effcond}
\end{equation}
where $\langle\phi_m,G\phi_n\rangle$ are the matrix elements of the coupling operator $G$ \eqref{n2} in the system energy basis. 

	\item[(b)] Simplicity of resonance energies. We assume that the so-called {\em level shift operators} $\Lambda_e$, $e\in{\mathcal E}_0=\{ E_m-E_n\ :\ m,n=1,\ldots, N\}$, have simple eigenvalues. The $\Lambda_e$ are explicit matrices describing the second order ($\lambda^2$) corrections to the energy differences $e\in{\mathcal E}_0$, given in Section \ref{FGRsect}.
\end{itemize}
\end{itemize}

The assumption (A1) on the form factor is more restrictive than $g\in L^2_{\rm obs}$ ({\em c.f.} point (a) at the beginning of Section \ref{sec:oa}) in that it requires a more stringent infrared behaviour (values of $p$) and the existence of one more derivative, relative to functions in $L^2_{\rm obs}$. The condition \eqref{effcond} ensures that the interaction does not suppress second order ($\lambda^2$) transition processes in S due to the coupling with R. The reservoir spectral density \eqref{specdens} governs these transitions since the coupling \eqref{n2} is linear in the field operator. This linear form of the interaction is not necessary for our method to work -- other interactions will lead to explicit conditions of effective coupling but are not expressed in terms of $J(\omega)$.  The assumptions (A1) together with (A2a) guarantee that the coupled system-reservoir complex has a unique stationary state, the coupled equilibrium state,  for small nonzero $\lambda$, see \cite{Merkli2001}.  (A2b) is a simplifying assumption that can be quite easily removed by a slightly more cumbersome analysis. 
\medskip

Here is our main result.

\begin{thm}[SR dynamics for correlated initial states]
	\label{theorem1}
	There is a constant $\lambda_0>0$ such that if $|\lambda|<\lambda_0$, then the following holds. 
	Let $\rho_{\s\r}$ be an initial system-reservoir state of the form \eqref{c62} and let $\rho_\s$ be its reduction to $\s$ \eqref{redin}. Then for all $t\ge 0$, $A\in\mathcal O$,
\begin{equation}
		\rho_{\s\r}\big(e^{\i t H_\lambda} A e^{-\i t H_\lambda}\big) = \big( e^{t{\mathcal L}_\s(\lambda)} \rho_\s\otimes\omega_{\r,\beta} \big)(A) + \chi(\lambda,t,A) + R(\lambda,t,A),
		\label{1.20}
\end{equation}
	where  ${\mathcal L}_\s(\lambda)$ is the Davies generator \eqref{daviesgen} and the remainder $R(\lambda,t,A)$ satisfies
	\begin{equation}
	| R(\lambda,t, A)| \le C(A)|\lambda|^{1/4}.
	\label{1.21-1}
\end{equation}
The dispersive term satisfies
\begin{eqnarray}
\chi(\lambda,t,A)&=&0 \qquad \mbox{if\ \  $\rho_{\s\r}=\rho_\s\otimes\omega_{\r,\beta}$}  \label{3.26}\\	
\chi(\lambda,t, A_\s\otimes\bbbone_\r) & =&0 \label{1.24-1}\\
|\chi(\lambda,t, A)| &\le& \frac{C(A)}{1+t^3}, \label{1.23}
\end{eqnarray}
and moreover,  
\begin{equation}
|\chi(\lambda,t,A)-  \big(\rho_{\s\r}- \rho_\s\otimes\omega_{\r,\beta}\big) (e^{\i t H_0}  A e^{-\i t H_0} )| \le C(A)\, |\lambda|.
\label{unifbound}
\end{equation}
The constants $C(A)$ depend on $A$ as well as the initial state $\rho_{\s\r}$ (but are independent of $t,\lambda$). For  $A=A_\s\otimes\bbbone_\r$ we have  $C(A_\s\otimes\bbbone_\r)\le c\|A_\s\|$ with $c$ independent of $A_\s$.
\end{thm}

\subsection{Implications}

Theorem \ref{theorem1} has the following consequences.

\begin{itemize}
\item[ {\bf (1)}] {\bf Validity of the Born approximation for initially factorized states} 

For initial states $\rho_{\s\r}=\rho_\s\otimes\omega_{\r,\beta}$, where $\rho_\s$ is any system density matrix, and for all $t\ge 0$, $A\in\mathcal O$, we have
\begin{equation}
	\rho_{\s\r}\big(e^{\i t H_\lambda} A e^{-\i t H_\lambda}\big) = \big( e^{t{\mathcal L}_\s(\lambda)} \rho_\s\otimes\omega_{\r,\beta} \big)(A) + O(|\lambda|^{1/4}).
	\label{1.20-0}
\end{equation}
This shows that if the system and reservoir are initially in a product state with the reservoir in equilibrium, then the total state stays (up to $ O(|\lambda|^{1/4})$) of product form for all times,  and the reservoir stays in its equilibrium state.

\item[ {\bf (2)}] {\bf Validity of the Markov approximation for initially correlated states} 

Taking in \eqref{1.20} observables of the form $A=A_\s\otimes\bbbone_\r$ shows the validity of the Markovian approximation, \eqref{14-i}, {\em even for initially correlated $\s\r$ states}. Namely, denoting by $\rho_\s^t$ the reduction of the full state $\rho_{\s\r}=\rho_{\s\r}( e^{\i t H_\lambda} \cdot e^{-\i t H_\lambda})$ at time $t$, and $\rho_\s=\rho_\s^{t=0}$, we have 
\begin{equation}
\sup_{t\ge 0}\big\|\rho_\s^t - e^{t{\mathcal L}_\s(\lambda)}\rho_\s\big\| \le C|\lambda|^{1/4},
\label{1.24}
\end{equation}
for a constant $C$ independent of $\lambda$, but generally depending on the initial state $\rho_{\s\r}$. (As $\rho_\s^t$ is a finite-dimensional density matrix, it is not necessary to specify which norm we take in \eqref{1.24}. We may take the trace norm.)

\item[ {\bf (3)}] {\bf Decay of correlations and emergence of the Born approximation regime} 

The term $\chi$ in \eqref{1.20} describes the evolution of the correlation between the system and reservoir. According to \eqref{1.23} it decays as $t^{-3}$. At $t=0$, \eqref{1.20} gives $\chi(\lambda,t=0,A)
= \rho_{\s\r}(A)-{\rm tr}_\r\{\rho_{\s\r}\} \otimes\omega_{\r,\beta}+ O(|\lambda|^{1/4})$, so if the initial state $\rho_{\s\r}$ differs from ${\rm tr}_\r\{\rho_{\s\r}\}\otimes\omega_{\r,\beta}$ then for small times, the correlation term $\chi(\lambda,t,A)$ can dominate the Markovian term $(e^{t{\mathcal L}_\s(\lambda)}\rho_\s\otimes\omega_{\r,\beta})(A)$ in \eqref{1.20}. Whether, and by how much, this happens depends on the observable $A$ -- for example, as explained in the previous point, for $A=A_\s\otimes\bbbone_\r$ we  have $\chi(\lambda,t,A)=0$.   To see which one, the Markovian or the correlation contribution, is dominant, we use the spectral decomposition \cite{MAOP}
\begin{equation}
\label{2.40}
e^{t{\mathcal L}_\s(\lambda)} =\sum_{j} e^{\i t (e_j +\lambda^2 a_j)} P_j,
\end{equation}
where the $P_j$ are spectral projections, the $e_j\in\mathbb R$ are the Bohr energies of the system (differences of eigenvalues of $H_\s$) and $a_j\in\mathbb C$, ${\rm Im}a_j\ge 0$ are the complex energy corrections (resonances) induced by the interaction with the reservoir. The Markovian dynamics \eqref{2.40}  exhibits exponential time decay  $\sim e^{- \lambda^2\gamma t}$  in all directions $P_j$ but one, which represents the projection onto the system equilibrium state at temperature $\beta$ (and for which $e_j=a_j=0$). The decay (approach to equilibrium) is exponential at a rate $\propto \lambda^2$, which is very slow for small coupling $\lambda$. In contrast, the polynomial decay of the correlation term, $t^{-3}$, happens at a rate which is independent of $\lambda$, meaning that the constant in \eqref{1.23} does not depend on $\lambda$. (Indeed, that constant encodes the fact that the reservoir dynamics alone is dispersive away from the projection onto the reservoir equilibrium state.)  It follows that for small coupling, the correlation term $\chi$ can be  dominant for small times, but as time increases, the Markovian term becomes dominant. Then a long time later ($\lambda$ small), the exponentially decaying functions lie below the power decay function, and the correlation term is leading once again, even though by that time, both those terms are smaller than the remainder $ O(|\lambda|^4)$.

If the interacting $\s\r$ complex is left to its own devices (no external influence) for sufficiently long, then according to \eqref{1.20}, and since $\e^{t \mathcal L(\lambda)}\rho_\s$ converges to $\rho_{\s,\beta}\propto e^{-\beta H_\s}$, the $\s\r$ state is approximately equal to the time-independent (uncoupled equilibrium) state $\rho_{\s,\beta}\otimes\omega_{\r,\beta}$. One may see this as an {\em a posteriori} justification for considering product initial states in certain circumstances. 
 \end{itemize}

\noindent
Further remarks.

\begin{itemize}
\item[(a)] The power of the decay in \eqref{1.23} depends on the smoothness of the form factor $g$, \eqref{n2}, which in turn determines the decay speed of the reservoir correlation function \cite{Markov1, Markov2}. We get higher powers $\sim t^{-n}$ assuming that $g$ is $(n+1)$ times differentiable and that the test functions in $L^2_{\rm obs}$ are $n$ times differentiable. Under a certain analyticity condition on $g$ (see \cite{MAOP}) and on the test functions in $L^2_{\rm obs}$, one can show that the time decay in \eqref{1.23} is exponential, with a $\lambda$-independent decay rate.

\item[(b)] We cannot ascertain in general that the constants $C(A)$  in Theorem \ref{theorem1} are bounded in the values of $t>0$ for time-dependent observables of the form $A=e^{\i t H_0}A'e^{-\i t H_0}$. This is so since the constants will depend on Sobolev norms of test functions appearing in the observables, that is, on their smoothness (the $L^2({\mathbb R}^3,d^3k)$ norm of their derivatives w.r.t. $|k|$). But $e^{\i t H_\r} W(f)e^{-\i t H_\r} = W(e^{\i |k|t}f)$ has a test function $e^{\i |k|t}f(k)$ that is rough (large $|k|$-derivative) for large $t$. Nevertheless, if $A=A_\s\otimes\bbbone_\r$, then $C(e^{\i t H_\s}A_\s e^{-\i H_\s t}\otimes\bbbone_\r)\le c\|A_\s\|$ is bounded uniformly in time.

\item[(c)] In the weak coupling regime one takes simultaneously $\lambda\rightarrow 0$ and $t\rightarrow\infty$, keeping $\tau=\lambda^2 t\in\mathbb R$ fixed. By \eqref{1.21-1}, \eqref{1.23} and the previous discussion point, 
\begin{eqnarray}
\lim_{\lambda\rightarrow 0} R\Big(\lambda,\tfrac{\tau}{\lambda^2},(e^{-\i\frac{\tau}{\lambda^2}H_\s}\otimes\bbbone_\r) A (e^{\i\frac{\tau}{\lambda^2}H_\s}\otimes\bbbone_\r)\Big) &=&0\nonumber\\
\lim_{\lambda\rightarrow 0} (\rho_{\s\r}- \rho_\s\otimes\omega_{\r,\beta}) \Big(e^{\i \frac{\tau}{\lambda^2} H_0}  (e^{-\i\frac{\tau}{\lambda^2}H_\s}\otimes\bbbone_\r) A (e^{\i\frac{\tau}{\lambda^2}H_\s}\otimes\bbbone_\r) e^{-\i \frac{\tau}{\lambda^2} H_0}\Big)&=&0.
\nonumber
\end{eqnarray}
Thus \eqref{1.20}  gives for all $\tau>0$, $A\in\mathcal O$
\begin{equation*}
	\lim_{\lambda\rightarrow 0}\  \rho_{\s\r}\Big(e^{\i \frac{\tau}{\lambda^2} H_\lambda} (e^{-\i\frac{\tau}{\lambda^2}H_\s}\otimes\bbbone_\r) A (e^{\i\frac{\tau}{\lambda^2}H_\s}\otimes\bbbone_\r)e^{-\i \frac{\tau}{\lambda^2} H_\lambda}\Big) = \big(e^{\tau \mathcal K } \rho_\s\otimes\omega_{\r,\beta} \big)(A).
\end{equation*}
This shows that in the weak coupling scaling limit, the total system-reservoir state (in the interaction picture when the system dynamics is removed) is of the product form $(e^{\tau \mathcal K } \rho_\s)\otimes\omega_{\r,\beta}$ for all $\tau>0$. This fact was already observed in \cite{Tasakietal}.

\item[(d)] If $A$ belongs to the norm closure of ${\mathcal O}$, then take an arbitrary $\epsilon>0$ and $A_\epsilon \in\mathcal O$ such that $\|A-A_\epsilon\|<\epsilon$. The quantity $\rho_{\s\r}(e^{\i t H_\lambda}Ae^{-\i t H_\lambda})$ is still well defined and we have $\rho_{\s\r}(e^{\i t H_\lambda} A e^{-\i t H_\lambda}) - \rho_{\s\r}(e^{\i t H_\lambda}A_\epsilon e^{-\i t H_\lambda}) = O(\epsilon)$, uniformly in $t$. Theorem \eqref{theorem1} then gives the expansion \eqref{1.20} for $\rho_{\s\r}(e^{\i t H_\lambda}A_\epsilon e^{-\i t H_\lambda})$, with constants $C(A_\epsilon)$ depending on $\epsilon$.  In the main term,  $\big( e^{t{\mathcal L}_\s(\lambda)} \rho_\s\otimes\omega_{\r,\beta} \big)(A_\epsilon) + \chi(\lambda,t,A_\epsilon)$, we can replace $A_\epsilon$ again by $A$, making an error of $O(\epsilon)$, uniformly in $\lambda$ and $t$ (see also \eqref{c31-1}). This shows that \eqref{1.20} remains valid for any $A$ in the norm closure of $\mathcal O$, modulo adding an arbitrarily small term $O(\epsilon)$ and allowing the constants $C(A)$ to depend also on $\epsilon$. One may use a similar argument to extend the result of Theorem \ref{theorem1} to $\nu=\infty$ in \eqref{c62}.

\end{itemize}
\medskip

\noindent
{\bf Connections.} The mathematical method we are using is based on \cite{Markov1,Markov2}. To our knowledge, this is the only approach able to handle perturbation theory in small $\lambda$ valid for all times $t\ge 0$. On the physics side, our work is close to \cite{Tasakietal, Yuasaetal}, where the authors show that the master equation is valid for correlated initial states in the weak coupling limit (see the Remark (c) above). Their technique is based on the Nakajima-Zwanzig projection method and does not extend beyond the weak coupling scaling regime.  In some aspects, the setup in \cite{Tasakietal, Yuasaetal}  is more general (allowing to treat NESS for example -- an extension of our methods to include this is planned) and in some aspects it is less general (their Kraus operators have to be bounded for the main result in \cite{Tasakietal} -- even though unbounded ones are used in applications given in \cite{Yuasaetal}). 

One of our main conclusions is that initial correlations do not invalidate the Markovian approximation. This means that the distinguishability (trace distance) of reduced system states cannot increase  during the evolution (by more than $O(|\lambda|^{1/4})$). An important feature responsible for this is that the reservoir dynamics is dispersive, meaning that the evolution converges to the stationary (equilibrium) state. This is built into the model by taking the infinite volume limit (continuous reservoir mode frequencies) and by the continuity of the spectrum of $H_\r$ (see also \cite{Markov1, Markov2, MAOP}). In contrast, if $\r$ is {\em finite}, an $M$-level system with discrete energy levels, then different effects appear. Those are studied in a variety of papers. It is shown in \cite{Laineetal, Devietal} that for finite-dimensional $\s$ and $\r$, the backflow of information $\r\rightarrow \s$ can lead to an {\em increase} in the  distinguishability if the $\s\r$ state is initially correlated, an effect which cannot happen for initial product states. The increase is a measure for non-Markovianity of the dynamics. In this situation, initial correlations lead to non-Markovianity of the system dynamics. A related question is whether and how the system dynamics can be represented by completely positive,  Kraus representation maps.  In \cite{Hayashietal}, the authors consider finite-dimensional $\s$ and $\r$ and show that a Kraus representation is valid for all times and for general initial correlations if and only if the joint dynamics is local unitary (no $\s\r$ coupling). It is shown in \cite{Pazetal} that the dynamics of an open $N$-level system can be described by at most $N^2$ completely positive trace preserving maps. The result is built on a decomposition of the (correlated) initial state using a so-called bath-positive decomposition.

In \cite{Alipouretal}, a new correlation picture approach is used to show that the reduced system dynamics satisfies a Lindblad-like master equation, in which the jump operators depend on the initial $\s\r$ correlation, yielding a nonlinear evolution equation. The philosophy there is to build a weak correlation, rather than a weak coupling, perturbation theory.  The corresponding perturbation series does not converge uniformly in time, but allows to analyze a strongly interacting $\s\r$ dynamics for finite times.

\section{Proof of Theorem \ref{theorem1}}

\subsection{Purification of $\rho_{\s\r}$}
\label{purifsect}

Our approach follows the works \cite{Markov1, Markov2} (see also \cite{KM1}) and we adopt the notation of \cite{Markov2}. The first step is to work with a  purification of the initial state $\rho_{\s\r}$. Consider   $\rho_\s\otimes\omega_{\r,\beta}$, 
where $\rho_\s$ is an arbitrary density matrix acting on ${\mathbb C}^N$ and $\omega_{\r,\beta}$ is the reservoir equilibrium state  \eqref{1.10}. The state $\rho_\s\otimes\omega_{\r,\beta}$ is represented by a {\em vector} 
\begin{equation}
	\Psi_0=\Psi_\s\otimes\Omega_\r \in {\mathcal H}_\gns
	\label{n60}
\end{equation}
in the new `purification' Hilbert space
\begin{equation}
{\mathcal H}_\gns  = {\mathcal H}_\s\otimes{\mathcal H}_\r
\label{3.2}
\end{equation}
where
\begin{equation}
	{\mathcal H}_\s = {\mathbb C}^N\otimes {\mathbb C}^N\quad 
\mbox{and}\quad 
{\mathcal H}_\r = \bigoplus_{n\ge 0} L^2_{\rm sym}\big( ({\mathbb R}\times S^2)^n, (du\times d\Sigma)^n\big)
\label{n48}
\end{equation}
is the (symmetric) Fock space over the one-particle space 
\begin{equation}
L^2( {\mathbb R}\times S^2, du\times d\Sigma)\equiv L^2( {\mathbb R}\times S^2).
\label{2.7}
\end{equation} 
Here, $d\Sigma$ is the uniform measure over the unit sphere $S^2\subset {\mathbb R}^3$. The link between $\rho_\s\otimes\omega_{\r,\beta}$ and \eqref{n60} is given by a representation map 
\begin{equation}
	\pi=\pi_\s\otimes\pi_\r: {\mathcal O}\rightarrow {\mathcal L}({\mathcal H}),
	\label{2.17}
\end{equation} 
mapping observables $A$ acting on ${\mathbb C}^N\otimes{\mathcal F}$ (before purification \eqref{2.2}) to operators on ${\mathcal H}_\gns$ (after purification \eqref{3.2}), in such a way that 
\begin{equation}
	\big(\rho_\s\otimes\omega_{\r,\beta}\big)( A) = \langle \Psi_0, \pi( A)\Psi_0\rangle_{{\mathcal H}_\gns}. 
	\label{n62}
\end{equation}
The vector $\Omega_\r$ in \eqref{n60} is the vacuum vector in the Fock space ${\mathcal H}_\r$ \eqref{n48}. It is the purification of the reservoir thermal equilibrium. The vector $\Psi_\s\in{\mathbb C}^N\otimes{\mathbb C}^N$ is the purification of the system density matrix $\rho_\s$. For example, the purification of the system Gibbs state $\rho_{\s,\beta}$ \eqref{c61} is given by 
\begin{equation} 
 \Omega_{\s,\beta} = Z_{\s,\beta}^{-1/2} \sum_{j=1}^N e^{-\beta E_j/2}\phi_j\otimes\phi_j \in{\mathbb C}^N\otimes{\mathbb C}^N,
\label{n47.1}
\end{equation}
where $Z_{\s,\beta}={\rm tr}_\s\, e^{-\beta H_\s}$.  The representation map $\pi$, \eqref{2.17} is explicitly given by 
\begin{eqnarray}
 \pi_\s(A_\s) &=&A_\s\otimes\bbbone_\s,
\label{n44.2}\\
\pi_\r(W(f)) &=& W_\beta(\tau_\beta f),
\label{3.9}
\end{eqnarray} 
where
\begin{equation}
	\tau_\beta : L^2({\mathbb R}^3, d^3k)\rightarrow  L^2( {\mathbb R}\times S^2, du\times d\Sigma)
\end{equation}
takes a function $f(k)$, $k\in{\mathbb R}^3$, into the function $(\tau_\beta f)(u,\Sigma)$, $u\in\mathbb R$, $\Sigma\in S^2$, defined by
\begin{equation}
	\big(\tau_\beta f\big)(u,\Sigma) = \sqrt{\frac{u}{1-\e^{-\beta u}}} \   |u|^{1/2}\left\{
	\begin{array}{ll}
		f(u,\Sigma), & u\ge0\\
		-\bar f(-u,\Sigma) & u<0
	\end{array}
	\right. .
	\label{n50}
\end{equation}
On the right side of \eqref{n50}, $g$ is represented in spherical coordinates, $u=|k|\ge 0$, $\Sigma\in S^2$. The Weyl operator $W(f)$ on the left side of \eqref{3.9} is given by \eqref{1.8} and is defined for $f\in L^2({\mathbb R}^3, d^3k)$. The Weyl operator on the right side is defined for functions $f\in L^2( {\mathbb R}\times S^2, du\times d\Sigma)$, as $W_\beta(f) =e^{\i \varphi_\beta(f)}$, where $\varphi_\beta(f)=\frac{1}{\sqrt 2}[a_\beta^*(f) +a_\beta(f)]$ is the field operator in the Fock space ${\mathcal H}_\r$, \eqref{n48}, smoothed out with $f$. The representation map $\pi_\r$ is also defined on creation operators, by the explicit formula
\begin{equation}
	\pi_\r(a^*(f)) = a_\beta^*\Big(\textstyle\sqrt{\frac{u}{1-\e^{-\beta u}}}\ |u|^{1/2} f(u,\Sigma)\chi_+(u) \Big) - a_\beta\Big(\textstyle\sqrt{\frac{u}{1-\e^{-\beta u}}}\ |u|^{1/2} \bar f(-u,\Sigma)\chi_-(u) \Big), 
\label{c45}
\end{equation}
where $\chi_+(u)=1$ if $u\ge 0$ and $\chi_+(u)=0$ for $u<0$ (and $\chi_-(u)=1-\chi_+(u)$). This is a convenient representation of the CCR, unitarily equivalent to the Araki-Woods thermal representation, and was introduced in \cite{JP}. We refer to \cite{BR,MAOP,Markov2,MLnotes} for additional detail. The purification of the initial state $\rho_{\s\r}$ \eqref{c62} is given by 
\begin{equation}
\sum_{\alpha=1}^\nu	\rho_{\s,\beta}\otimes\omega_{\r,\beta}\big(K_\alpha^*   A K_\alpha\big) =\sum_{\alpha=1}^\nu	 \langle\Omega_{\s\r,\beta,0}, \pi(K_\alpha^*) \pi(A) \pi(K_\alpha)\Omega_{\s\r,\beta,0}\rangle,
	\label{n71-1}
\end{equation}
where the inner product is that of ${\mathcal H}_\gns$ \eqref{3.2} and
\begin{equation}
	\Omega_{\s\r,\beta,0} = \Omega_{\s,\beta}\otimes\Omega_\r
	\label{c7}
\end{equation}
is the uncoupled equilibrium (KMS) state. We note that $\pi$ is well defined on polynomials in creation and annihilation operators (and limits thereof), see \eqref{2pt}, and $\pi({\mathfrak C})$ is a well defined set of unbounded operators on ${\mathcal H}_\gns$.  We prove in Lemma \ref{lemma4.4} below that $\pi(K)\Omega_{\s\r,\beta,0}$ is well defined for any $K\in\mathfrak C$. This guarantees that \eqref{n71-1} is well defined. One can see the right hand side of \eqref{n71-1} as the {\em definition} of our state $\rho_{\s\r}$.

\subsection{Coupled dynamics and Liouville operator}
\label{s2.2}

The uncoupled Heisenberg dynamics $\alpha^t_0(A) = e^{\i t H_0}  A e^{-\i t H_0}$ leaves $\mathcal O$ invariant (see \eqref{1.14})  but the interacting dynamics $e^{\i t H_\lambda} Ae^{-\i t H_\lambda}$ does not. It is thus not {\em a priori} clear how to represent $e^{\i t H_\lambda} Ae^{-\i t H_\lambda}$ as an operator in ${\mathcal H}_\gns$. However, we can extend the domain of $\pi$ and define a self-adjoint {\em Liouville operator} on $\mathcal H_\gns$, such that for $A\in\mathcal O$,
\begin{equation}
	\pi\big( e^{\i t H_\lambda}  
	A e^{-\i t H_\lambda}\big) = e^{\i t L_\lambda}\pi(A)e^{-\i t L_\lambda}.
	\label{n63}
\end{equation}
To do this we proceed in a standard way using the Dyson expansion 
\begin{equation}
e^{\i t H_\lambda} Ae^{-\i t H_\lambda} = \alpha^t_0(A) +\sum_{n\ge 1} (\i \lambda)^n \int_{0\le t_1\le\cdots\le t_n\le t}  dt_1\cdots dt_n \ T_{t_1,\ldots,t_n}(A),
\label{4.15}
\end{equation}
where
\begin{equation} T_{t_1,\ldots,t_n}(A)= [\alpha^{t_n}_0(V),[\alpha_0^{t_{n-1}}(V),\ldots,[\alpha^{t_1}_0(V), \alpha^t_0(A)]\ldots]]
\end{equation}
is the multiple commutator and $V=G\otimes\varphi(g)$ is the interaction operator \eqref{n2}. The series \eqref{4.15} converges in the strong sense on suitable states. It is easy to check that 
\begin{equation}
\pi(e^{\i t H_0} A e^{-\i t H_0}) = e^{\i t L_0}\pi(A)e^{-\i t L_0},
\end{equation}
where the uncoupled Liouville operator is (not writing obvious factors $\bbbone$)
\begin{equation}
	L_0 = L_\s +L_\r,\quad 
	L_\s = H_\s\otimes\bbbone_\s\ - \bbbone_\s\otimes H_\s, \quad 
	L_\r =  d\Gamma(u).
	\label{4}
\end{equation}
Here,  $d\Gamma(u)$ is the second quantization of multiplication by the radial variable $u\in{\mathbb R}$, acting on the Fock space ${\mathcal H}_\r$ \eqref{n48}. In particular, $\e^{\i t L_\r} a_\beta(f)e^{-\i t L_\r} = a_\beta(e^{\i t u}f)$ is a Bogoliubov transformation. We now apply $\pi$ to the right side of \eqref{4.15}. More precisely, one shows that $\pi$ applied to the truncated series $\sum_{n=1}^N\ldots$ has a strong limit as $N\rightarrow \infty$. Due to the structure of the sum, the limit operator has again the form of a Dyson series and equals $e^{\i t L_\lambda}\pi(A)e^{-\i t L_\lambda}$ (see for instance Section 2.1.3 of \cite{FM-TI} for details). The self-adjoint operator $L_\lambda$ is called the {\em Liouville operator} and has the form
\begin{equation}
	L_\lambda = L_0 +\lambda I,
	\label{n66}
\end{equation}
where the interaction operator $I$ in \eqref{n66} is given by $I = \pi\big(G\otimes \varphi(g)\big) - J\pi\big(G\otimes \varphi(g)\big)J$ (compare with \eqref{n2}), where $J$ is the modular conjugation. The explicit form of $J$ is well known, see {\em e.g.} (1.21) of \cite{KoMeSo}. One then gets the expression  \cite{KoMeSo,MAOP}
\begin{equation}
	I = G\otimes \bbbone_\s\otimes \varphi_\beta(\tau_\beta g) - \bbbone_\s\otimes {\mathcal C} G{\mathcal C}\otimes \varphi_\beta(e^{-\beta u/2} \tau_\beta g),
	\label{n40}
\end{equation}
where $\mathcal C$ is the operator taking complex conjugation of components of vectors in ${\mathbb C}^N$ written in the eigenbasis of $H_\s$.
The construction of $L_\lambda$ associated to $H_\lambda$ is well understood \cite{JP,BFS,DJP,MAOP}.  The upshot is that the right side of \eqref{n63} defines the dynamics of the infinitely extended system.

From perturbation theory of equilibrium (KMS) states \cite{DJP,BR} we know that the interacting system-reservoir complex has a unique equilibrium state, represented by a normalized vector $\Omega_{\s\r,\beta,\lambda}\in{\mathcal H}$, satisfying
\begin{equation}
	L_\lambda \Omega_{\s\r,\beta,\lambda}=0.
	\label{n66.2}
\end{equation}
The vector $\Omega_{\s\r,\beta,\lambda}$  is analytic in $\lambda$ at the origin, with (c.f. \eqref{c7})
\begin{equation}
\| \Omega_{\s\r,\beta,\lambda} -\Omega_{\s\r,\beta,0}\| \le C |\lambda|.
	\label{c12}
\end{equation}

\subsection{Resonance expansion and proof of Theorem \ref{theorem1}}
\label{s2.3}

The expectation of a system-reservoir observable $A\in\mathcal O$ in the state $\rho_{\s\r}$ at time $t$ is 
\begin{equation}
\rho_{\s\r}(e^{\i t H_\lambda} A e^{-\i t H_\lambda}) = \sum_{\alpha=1}^\nu \rho_{\s,\beta}\otimes\omega_{\r,\beta}\big(K^*_\alpha e^{\i t H_\lambda}  A e^{-\i t H_\lambda} K_\alpha\big).
\label{c6}
\end{equation}
We now analyze one of the terms in the sum, for a fixed $\alpha$,  writing simply $K$ for $K_\alpha$, and we will restore the sum over the $\alpha$ at the end.  According to \eqref{n71-1}, \eqref{n63} we have
\begin{equation}
	\rho_{\s,\beta}\otimes\omega_{\r,\beta}\big(K^* \e^{\i t H_\lambda} A \e^{-\i t H_\lambda}K\big) = \langle\Omega_{\s\r,\beta,0}, \pi(K^*) e^{\i t L_\lambda}\pi(A)e^{-\i tL_\lambda} \pi(K)\Omega_{\s\r,\beta,0}\rangle,
	\label{n71}
\end{equation}
where $\Omega_{\s\r,\beta,0}$ is the uncoupled KMS state \eqref{c7}.  The representation $\pi'(\cdot) = J\pi(\cdot)J$ has the property that $\pi'(X)$ and $\pi(Y)$ commute for all operators $X,Y$ (Tomita-Takesaki, \cite{BR,MAOP}); for unbounded $X,Y$ the commutation is understood in the strong (or weak) sense on suitable vectors. We now use the identity
\begin{equation}
\pi(X)\Omega_{\s\r,\beta,0} = \pi'\big( e^{-\beta H_0/2} X^* e^{\beta H_0/2}\big)\Omega_{\s\r,\beta,0},
\label{c8}
\end{equation}
which holds provided the operator $e^{-\beta H_0/2} X^* e^{\beta H_0/2}$ is well defined (note that $e^{\beta H_0/2}$ is unbounded). 
The relation \eqref{c8} is derived as follows. We have
\begin{eqnarray}
\pi(X)\Omega_{\s\r,\beta,0} &=& J\Delta^{1/2} \pi(X^*)J\Delta^{1/2}\Omega_{\s\r,\beta,0}\nonumber\\
&=& J\Delta^{1/2}\pi(X^*)\Delta^{-1/2}J\Omega_{\s\r,\beta,0}.
\label{c9}
\end{eqnarray}
The first equality in \eqref{c9} is the defining property of $J$ and $\Delta$ and the modular operator is $\Delta=e^{-\beta L_0}$. The second equality in \eqref{c9} follows from $JL_0=-L_0J$ \cite{BR,DJP,BFS,MAOP}. Next, 
\begin{eqnarray}
\Delta^{1/2}\pi(X^*)\Delta^{-1/2} &=& e^{\i t L_0}\pi(X^*)e^{-\i t L_0}\big|_{t=\i \beta/2}\nonumber\\
& =& \pi(e^{\i t H_0} X^* e^{-\i t H_0})\big|_{t=\i \beta/2} \nonumber\\
&=& \pi(e^{-\beta H_0/2} X^* e^{\beta H_0/2}).
\label{c10}
\end{eqnarray}
Using that $J\pi(\cdot)J=\pi'(\cdot)$ and combining \eqref{c10} with \eqref{c9} yields \eqref{c8}. Due to \eqref{c8} we may replace, in \eqref{n71}, $\pi(K)$ by $\pi'(e^{-\beta H_0/2} K^* e^{\beta H_0/2})$ and use that the latter operator commutes with $e^{\i t L_\lambda} \pi(A)e^{-\i t L_\lambda}$,
\begin{eqnarray}
\lefteqn{
\rho_{\s,\beta}\otimes\omega_{\r,\beta}\big(K^* \e^{\i t H_\lambda} A \e^{-\i t H_\lambda}K\big)}\nonumber\\
&& = \big\langle\Omega_{\s\r,\beta,0}, \pi(K^*) \pi'(e^{-\beta H_0/2} K^* e^{\beta H_0/2})e^{\i t L_\lambda}\pi(A)e^{-\i tL_\lambda} \Omega_{\s\r,\beta,0}\big\rangle.
\label{c11}
\end{eqnarray}
By \eqref{n66.2} we have $e^{-\i t L_\lambda}\Omega_{\s\r,\beta,\lambda} =\Omega_{\s\r,\beta,\lambda}$ and so
\begin{equation}
e^{\i t L_\lambda}\pi(A)e^{-\i t L_\lambda}\Omega_{\s\r,\beta,0} = e^{\i t L_\lambda}\pi(A) \Omega_{\s\r,\beta,\lambda}+e^{\i t L_\lambda}\pi(A)\big(  \Omega_{\s\r,\beta,0} - \Omega_{\s\r,\beta,\lambda}\big),
\label{c2.41}
\end{equation}
where the uncoupled equilibrium state $\Omega_{\s\r,\beta,0}$ is given in \eqref{c7}.  We combine \eqref{c11},  \eqref{c2.41}  and \eqref{c12} into
\begin{equation}
\rho_{\s,\beta}\otimes\omega_{\r,\beta}\big(K^* \e^{\i t H_\lambda} A\e^{-\i t H_\lambda}K\big) =\langle\phi, e^{\i t L_\lambda}\psi\rangle +R_1(t,\lambda),
\label{c13}
\end{equation}
where
\begin{eqnarray}
\phi&=&\pi(K)\pi'(e^{\beta H_0/2} K e^{-\beta H_0/2})\Omega_{\s\r,\beta,0}\nonumber\\
\psi &=& \pi(A)\Omega_{\s\r,\beta,0}
\label{c15}
\end{eqnarray}
and
\begin{equation}
|R_1(\lambda,t)| \le C|\lambda|\, \|A\|\, \big\| \pi(K)\pi'(e^{\beta H_0/2} K e^{-\beta H_0/2})\Omega_{\s\r,\beta,0}\big\|,
\label{c14}
\end{equation}
uniformly in $t\in\mathbb R$. Now we apply the results of \cite{Markov1} to the main term on the right side of \eqref{c13}. That work gives an expansion of $\langle \phi, e^{\i t L_\lambda}\psi\rangle$  for vectors $\phi$, $\psi$ satisfying 
\begin{equation}
	\phi,\psi\in{\mathcal D},\quad 	\bar L_\lambda \phi, \bar L_\lambda\psi\in{\mathcal D},
	\label{c16.1}
\end{equation}
where $\bar L_\lambda = P_\r^\perp L_\lambda P_\r^\perp|_{{\rm Ran}P_\r^\perp}$ and  where $\mathcal D\subset\mathcal H_\gns$ is a suitable dense set which we present in \eqref{n67.1} below. The orthogonal projection $P_\r$ is defined as
\begin{equation}
	P_\r= \bbbone_\s \otimes |\Omega_\r\rangle\langle\Omega_\r|, \qquad P_\r^\perp =\bbbone - P_\r.
	\label{3.34}
\end{equation} 
We show that \eqref{c16.1} holds in Section \ref{estKsect} and continue here the analysis using the result of Theorem 2.1 of \cite{Markov1},
\begin{equation}
	\langle\phi, e^{\i tL_\lambda}\psi\rangle  = \langle\phi, (e^{\i t M(\lambda)}\otimes P_\r)\psi\rangle +\langle \phi, P^\perp_\r e^{\i t P^\perp_\r L_\lambda P^\perp_\r} P^\perp_\r \psi\rangle  +R(\lambda,t),
	\label{c16}
\end{equation}
with a remainder satisfying
\begin{equation}
\big|R(\lambda,t)\big|\le C |\lambda|^{1/4} \, {\mathcal A}(\phi,\psi),
\label{c17}
\end{equation}
for constants $C$ and $\mathcal A$ independent of $\lambda$ and  $t\ge0$. 
 We now explain the operator $M(\lambda)$ in \eqref{c16}. Denote by ${\mathcal E}_0$ the set of all differences of eigenvalues of $H_\s$ ({\em i.e.}, the spectrum of $L_\s$, \eqref{4}). Denote the eigenprojection of $L_0$ associated to $e\in{\mathcal E}_0$ by $P_e=P_{\s,e}\otimes P_\r$, where $P_{\s,e}$ is the eigenprojection of $L_\s$ and define the {\em level shift operator}
\begin{equation}
	\Lambda_e = -P_e I P^\perp _e(L_0-e+\i 0_+)^{-1}  P^\perp _e IP_e,
	\label{n98-1}
\end{equation}
where $\i  0_+$ is the limit of $\i \epsilon$ as $\epsilon\rightarrow 0_+$ (the limit in \eqref{n98-1} exists in operator norm, see  \cite{Markov1,Markov2}). The operator 
$\Lambda_e$ acts on ${\rm ran}P_e$. It is explained in detail in \cite{Markov1} how $\Lambda_e$ determines the eigenvalues of $L_\lambda$ close to $e$.  Under the Fermi Golden Rule and simplicity conditions (A2a), (A2b), the operator $M(\lambda)$ in \eqref{c16} is given by
\begin{equation}
M(\lambda) = L_\s + \lambda^2 \Lambda,\qquad  \Lambda =\bigoplus_{e\in{\mathcal E}_0} \Lambda_e. 
\label{c21}
\end{equation}
The operators $L_\s$ and $\Lambda$ commute. 
Now that we have explained the objects defining the dynamics `along $P_\r$' in \eqref{c16} we can analyze it further. Writing 
\begin{equation}
\pi(A)= \sum_j \pi(A_\s^j \otimes A_\r^j) = \sum_j \pi_\s(A_\s^j) \otimes \pi_\r(A_\r^j)
\label{c18}
\end{equation}
and taking into account \eqref{c15} and \eqref{3.34},  it follows that 
\begin{eqnarray}
\langle\phi, (e^{\i t M(\lambda)}\otimes P_\r)\psi\rangle &=&
\sum_j \omega_{\r,\beta}(A_\r^j)\big \langle \phi, \big(e^{\i t M(\lambda)} \pi_\s(A_\s^j)\Omega_{\s,\beta}\big)\otimes \Omega_\r\big\rangle\nonumber\\
& =& \sum_j \omega_{\r,\beta}(A_\r^j)\big \langle \phi, \big( \pi_\s( e^{t{\mathcal L}_*} A_\s^j)\Omega_{\s,\beta}\big)\otimes \Omega_\r\big\rangle,
\label{c19}
\end{eqnarray}
where the operator ${\mathcal L}_*$ is uniquely defined by (see also \cite{MAOP, Markov1, Markov2})
\begin{equation}
e^{\i t M(\lambda)}\pi_\s(X)\Omega_{\s,\beta} = \pi_\s\big(e^{t {\mathcal L}_*}X\big)\Omega_{\s,\beta},\qquad \mbox{for all system operators $X\in\mathcal B({\mathbb C^N})$}.
\label{c24}
\end{equation}
The second factor of the summand of \eqref{c19} is ({\em c.f.} \eqref{c15})
\begin{eqnarray}
\lefteqn{
\big \langle \phi, \big(\pi_\s(e^{t{\mathcal L}_*} A_\s^j)\Omega_{\s,\beta}\big)\otimes \Omega_\r\big\rangle}\nonumber\\
&=&
\big\langle\pi(K)\Omega_{\s\r,\beta,0}, \pi\big(e^{t{\mathcal L}_*} A_\s^j\otimes\bbbone_\r\big)\pi'(e^{-\beta H_0/2} K^* e^{\beta H_0/2})\Omega_{\s\r,\beta,0}\big\rangle\nonumber\\
&=& \rho_{\s,\beta}\otimes\omega_{\r,\beta}\big(K^* (e^{t{\mathcal L}_*} A_\s^j\otimes\bbbone_\r)K \big),
\label{c25}
\end{eqnarray}
where we used \eqref{c6}, \eqref{c8} in the last step. Now we remember that we wrote $K$ for $K_\alpha$, and summing over $\alpha$, as per \eqref{c6}, we obtain from \eqref{c13}, \eqref{c16},
\begin{eqnarray}
\rho_{\s\r}(e^{\i t H_\lambda} Ae^{-\i t H_\lambda}) &=& \sum_\alpha \sum_j \omega_{\r,\beta}(A_\r^j)\,  \rho_{\s,\beta}\otimes\omega_{\r,\beta}\big(K_\alpha^* (e^{t{\mathcal L}_*} A_\s^j\otimes\bbbone_\r)K_\alpha \big)\nonumber\\
&&+ 
\big \langle \Phi, P^\perp_\r e^{\i t P^\perp_\r L_\lambda P^\perp_\r} P^\perp_\r \pi({ O}) \Omega_{\s\r,\beta,0} \big\rangle  +R_2(\lambda,t), 
\label{c27}
\end{eqnarray}
where
\begin{equation}
\Phi =\sum_{\alpha=1}^\nu \pi(K_\alpha)\pi'(e^{\beta H_0/2} K_\alpha e^{-\beta H_0/2})\Omega_{\s\r,\beta,0}
\label{c29}
\end{equation}
and the remainder is the sum of \eqref{c14} and \eqref{c17},
\begin{eqnarray}
|R_2(\lambda,t)| &\le& C|\lambda|^{1/4} \sum_{\alpha=1}^\nu {\mathcal A}\Big(\pi(K_\alpha)\pi'(e^{\beta H_0/2} K_\alpha e^{-\beta H_0/2})\Omega_{\s\r,\beta,0}\, ,\,  \pi(A)\Omega_{\s\r,\beta,0}\Big)\nonumber\\
&&+C|\lambda|\, \| A \|\, \sum_{\alpha=1}^\nu \big\| \pi(K_\alpha)\pi'(e^{\beta H_0/2} K_\alpha e^{-\beta H_0/2})\Omega_{\s\r,\beta,0}\big\|.
\label{c28}
\end{eqnarray}
In terms of the reduced initial state $\rho_\s$, defined in \eqref{redin},  the first term on the right side of \eqref{c27} is simply
\begin{eqnarray}
\lefteqn{
\sum_\alpha \sum_j \omega_{\r,\beta}( A_\r^j)\  \rho_{\s,\beta}\otimes\omega_{\r,\beta}\big(K_\alpha^* (e^{t{\mathcal L}_*} A_\s^j\otimes\bbbone_\r)K_\alpha \big)}\nonumber\\
&=& \sum_j \rho_\s\big( e^{t{\mathcal L}_*} A_\s^j\big) \omega_{\r,\beta} ( A_\r^j)\nonumber\\
&=& \sum_j \big( (e^{t{\mathcal L}}\rho_\s) \otimes \omega_{\r,\beta}\big) ( A_\s^j\otimes A_\r^j)=\big((e^{t{\mathcal L}} \rho_\s)\otimes\omega_{\r,\beta}\big) (A),
\label{c26}
\end{eqnarray}
where ${\mathcal L}$ is the adjoint of ${\mathcal L}_*$ with respect to the inner product $(X,Y) = {\rm tr}_\s X^*Y$ of the space of system operators. It is shown in the Appendix of \cite{MAOP} that $\mathcal L$ is the Davies generator. Combine \eqref{c26} and \eqref{c27},
\begin{equation}
\rho_{\s\r}(e^{\i t H_\lambda} Ae^{-\i t H_\lambda}) = \big((e^{t{\mathcal L}} \rho_\s)\otimes\omega_{\r,\beta}\big) ( A) +\chi(\lambda, t,A)+R_2(\lambda,t),
\label{c30}
\end{equation}
where (recall $\Phi$ is given in \eqref{c29})
\begin{equation}
\chi(\lambda, t,A)
 = \big \langle \Phi, P^\perp_\r e^{\i t P^\perp_\r L_\lambda P^\perp_\r} P^\perp_\r \pi(A) \Omega_{\s\r,\beta,0} \big\rangle.
 \label{c31-1} 
\end{equation}
This is the expansion \eqref{1.20} of Theorem \ref{theorem1}. The decay \eqref{1.23} follows from the regularity of the resolvent, as explained in (1.10) of \cite{Markov1} (see also the related Lemma \ref{lemma2.1} below).
Note that $P^\perp_\r\pi( A_\s\otimes\bbbone_\r)\Omega_{\s\r,\beta,0}=0$. Using this in \eqref{c31-1} shows \eqref{1.24-1}. Furthermore, for initial states $\rho_{\r\s}=\rho_\s\otimes\omega_{\r,\beta}$ the Kraus operators are of the form $K_\alpha=K'_\alpha\otimes\bbbone_\r$ and thus $P^\perp_\r\Phi=0$, see \eqref{c29}. Using this information in \eqref{c31-1} shows \eqref{3.26}.   We now prove \eqref{unifbound}. From \eqref{c31-1}, 
\begin{equation}
\chi(\lambda,t,A) = \langle \Phi, P_\r^\perp e^{\i t L_0}\pi(A)\Omega_{\s\r,\beta,0}\rangle +T(\lambda,t,A)
\label{2.55}
\end{equation}
where, writing $\bar X$ for $P_\r^\perp XP_\r^\perp|_{{\rm Ran}P_\r^\perp}$,
\begin{equation}
T(\lambda,t,A) = \i\lambda \int_0^t  \big\langle \Phi, P^\perp_\r e^{\i s \bar L_\lambda }\bar I e^{-\i(s-t)L_0} P^\perp_\r \pi(A) \Omega_{\s\r,\beta,0} \big\rangle ds.
\label{2.56}
\end{equation}
Using that $P^\perp_\r=\bbbone-P_\r$ one readily sees that 
\begin{equation}
\langle \Phi, P_\r^\perp e^{\i t L_0}\pi(A)\Omega_{\s\r,0}\rangle =  \rho_{\s\r}\big(e^{\i t H_0}  Ae^{-\i t H_0}\big) - (\rho_\s\otimes\omega_{\r,\beta})\big(e^{\i t H_0} Ae^{-\i t H_0}\big).
\label{2.57}
\end{equation}
In order to show the bound  \eqref{unifbound} it suffices to prove that for all $t\ge 0$, 
\begin{equation} 
|T(\lambda,t,A)|\le C|\lambda|.
\label{bnd}
\end{equation}
The clue in this bound is that the integrand  of \eqref{2.56} decays in $s$ sufficiently quickly to be integrable.  

As in \cite{Markov2} we define the norms $\|\cdot\|_j$,  for $j=0,1,2,\ldots$, on $\mathcal H_{\rm GNS}$ \eqref{3.2}: 
\begin{equation}
  \|\phi\|_j= \|(1+\bar \A^2)^{j/2}\phi\|, \qquad \A=d\Gamma(\i\partial_u),\qquad  \bar \A = P^\perp_\r \A P^\perp_\r|_{{\rm Ran}P_\r^\perp}.
  \label{norm0}
  \end{equation} 
We denote the $j$-fold commutator of $\bar X$ with $\bar \A$ by ${\rm ad}^j_{\bar \A}(\bar X)=[\cdots [[\bar X,\bar \A],\bar \A]\cdots]$. 

\begin{lem}
\label{lemma2.1}
Suppose the form factor $g$ satisfies $\|\partial_u^j\tau_\beta g\|_{L^2({\mathbb R}\times S^2)}<\infty$ for $j=0,\ldots,4$ and suppose $X$ is an operator such that $\|\N^{-1/2}{\rm ad}^j_{\bar \A}(\bar X)\N^{-1/2}\|<\infty$ for $j=1,2,3$, where $\N=d\Gamma(\bbbone_{\mathcal H_\r})$ is the number operator in the Fock space $\mathcal H_\r$ (see \eqref{n48} and also \eqref{N}). Then we have
\begin{equation}
\max_{1\le r\le 3}\Big|\big\langle \Phi
, (\bar L_\lambda-z)^{-1} \bar X (\bar L_0-\zeta)^{-r}\Psi\big\rangle\Big| \le C_{\bar X} \|\Phi\|_3\, \|\Psi\|_3,
\label{c2.58} 
\end{equation}
uniformly in $z,\zeta$ with ${\rm Im}z$, ${\rm Im}\zeta<0$. Here, $\|\cdot\|_j$ is the norm \eqref{norm0} and $C_{\bar X}$ is a constant depending on $\bar X$. 
\end{lem}

We give a proof of Lemma \ref{lemma2.1} below. For now we apply it to estimate the term \eqref{2.56}. We write the propagator $e^{-\i(s-t)L_0}$ in its resolvent representation (Fourier-Laplace transform, see also equation (1.30) in \cite{Markov1}), for $w>0$,
\begin{eqnarray}
	\lefteqn{
\big\langle \Phi, P^\perp_\r e^{\i s \bar L_\lambda }\bar I e^{-\i(s-t)L_0} P^\perp_\r \pi(A) \Omega_{\s\r,\beta,0} \big\rangle}\nonumber\\
&=&\frac{-1}{2\pi\i}\int_{{\mathbb R}-\i w}\frac{e^{\i(t-s)\zeta }}{(\zeta+\i)^2} \langle \Phi, P^\perp_\r e^{\i s \bar L_\lambda }\bar I (\bar L_0-\zeta)^{-1 }P^\perp_\r \Psi \rangle d\zeta,
\end{eqnarray}
where $\Psi = (\bar L_0+\i)^2\pi(A)\Omega_{\s\r,\beta,0}$. Next we proceed analogously for $e^{\i s\bar L_\lambda}$ to obtain (set $\tau=t-s$)
\begin{eqnarray}
T(\lambda,t,A)&=&\i\lambda \big(\frac{-1}{2\pi\i}\big)^2 \int_0^t d\tau \int_{\mathbb R-\i w}dz\frac{e^{\i (t-\tau)z}}{(z+\i)^2}\int_{{\mathbb R}-\i w}d\zeta\  \frac{e^{\i \tau \zeta}}{(\zeta+\i)^2} \nonumber\\
&& \times \langle\Phi', (\bar L_\lambda-z)^{-1} \bar I (\bar L_0-\zeta)^{-1}\Psi\rangle,
\label{2.60-1}
\end{eqnarray}
with $\Phi '= (\bar L_\lambda-\i)^2\Phi$. One shows that $\Psi$ and $\Phi'$ are well defined and $\|\Phi'\|_3, \|\Psi\|_3<\infty$ by the arguments of Section \ref{estKsect}. We divide the $\tau$ integral into $\tau\in[0,1]$ and $\tau\in[1,t]$, so 
\begin{equation}
T(\lambda,t,A) = T_1(\lambda,t,A)+ T_2(\lambda,t,A),
\end{equation}
where $T_1$ is given by \eqref{2.60-1} with $\int_0^t$ replaced by $\int_1^t$ and $|T_2(\lambda,t,A)|\le C|\lambda|$. To analyze $T_1$, we use (twice) that $e^{\i\tau\zeta} = \frac{1}{\i\tau}\partial_\zeta e^{\i\tau\zeta}$ and integrate by parts in $\zeta$, to arrive at
\begin{eqnarray}
T_1(\lambda,t,A)&=&\i\lambda \big(\frac{-1}{2\pi\i}\big)^2 \int_1^t d\tau \frac{-1}{\tau^2} \int_{\mathbb R-\i w}dz\frac{e^{\i(t-\tau) z}}{(z+\i)^2}\int_{{\mathbb R}-\i w}d\zeta\  e^{\i \tau\zeta}  \nonumber\\
&& \times\partial^2_\zeta\big\{ (\zeta+\i)^{-2} \langle\Phi', (\bar L_\lambda-z)^{-1} \bar I (\bar L_0-\zeta)^{-1}\Psi\rangle\big\}.
\label{c2.61}
\end{eqnarray}
We now use Lemma \ref{lemma2.1} to bound the term $\partial^2_\zeta\{ (\zeta+\i)^{-2} \langle\Phi', (\bar L_\lambda-z)^{-1} \bar I (\bar L_0-\zeta)^{-1}\Psi\rangle\}$, and we arrive at
\begin{equation}
|T_1(\lambda, t,A)|\le C |\lambda| \int_1^t \frac{d\tau}{\tau^2} \le C|\lambda|,\qquad \forall t\ge 1. 
\end{equation}
Note that in the application of Lemma \ref{lemma2.1}, we took $\bar X=\bar I$, and $\|{\mathcal N}^{-1/2}{\rm ad}_{\bar \A}^j (\bar I)\mathcal N^{-1/2}\|<\infty$ is guaranteed by the condition 
(A1) on the form factor $g$. This shows the relation \eqref{bnd} and hence \eqref{unifbound}. The proof of Theorem \ref{theorem1} is complete, modulo  a proof of Lemma \eqref{lemma2.1}, which we now present.



\medskip

{\em Proof of Lemma \ref{lemma2.1}. } As in the proof of Theorem 3.1 in \cite{Markov2}, one considers the regularization $\bar L_\lambda(\alpha) =\bar L_0 +\i \alpha \bar\N + \lambda \bar I(\alpha)$, $\alpha>0$, where
\begin{equation}
\bar I(\alpha) =\frac{1}{\sqrt{2\pi}}\int_{\mathbb R} \widehat f(s) e^{\i \alpha s \bar \A}\bar I e^{-\i \alpha s \bar \A} ds,\qquad  \A=d\Gamma(\i \partial_u)
\end{equation}
and $\widehat f$ is the Fourier transform of a Schwartz function $f$ satisfying $f^{(k)}(0)=1$, $k=0,1,2,\ldots$ (so $f(s)=e^s$ near $s=0$). The bound \eqref{c2.58} follows if the same ($\alpha$ independent) bound can be shown with $\bar L_\lambda, \bar L_0$ replaced by $\bar L_\lambda(\alpha), \bar L_0(\alpha)$, see Section 3.1.1 of \cite{Markov2}. Let us write for short $R_z=(\bar L_\lambda(\alpha)-z)^{-1}$ and $S_\zeta = (\bar L_0(\alpha)-\zeta)^{-1}$. We also write $X$ instead of $\bar X$ in the remainder of the proof, and $\A$ for $\bar \A$, $I$ for $\bar I$. Note that $S_\zeta = e^{\alpha \A}(L_0-\zeta)^{-1}e^{-\alpha \A}$. Using that $\partial_\alpha (S_\zeta)^k= \A S_\zeta^k -S_\zeta^k \A$ and $\partial_\alpha R_z = \A R_z-R_z\A+\lambda R_z Y R_z$, where
\begin{equation}
Y = \partial_\alpha I(\alpha) -[\A,I(\alpha)] = \frac{1}{\sqrt{2\pi}}\int_{\mathbb R}(\i s-1)\widehat f(s)  e^{\i \alpha s  \A}[\A, I] e^{-\i \alpha s  \A} ds,
\label{2.60}
\end{equation}
we get 
\begin{eqnarray}
\partial_\alpha  \langle \Phi
, R_z  X S_\zeta^3\Psi\rangle
&=& \langle \Phi
, \A R_z  X S_\zeta^3\Psi\rangle -\big\langle \Phi
, R_z X S_\zeta^3 \A \Psi\rangle\nonumber\\
&&+ \langle \Phi
, R_z[X,\A]  S_\zeta^3\Psi\big\rangle + \lambda \langle \Phi
, R_z YR_z  X S_\zeta^3\Psi\rangle.
\label{2.61}
\end{eqnarray}
By expanding $e^{\i \alpha s  \A}[\A, I] e^{-\i \alpha s  \A} $ in \eqref{2.60} in a power series in $\alpha$ and using that $f^{(k)}(0)=1$, one derives the bound $\|\N^{-1/2} Y\N^{-1/2}\|\le C\alpha^\ell$, for any $\ell=1,2,\ldots$, provided that  $\|\partial_u^j\tau_\beta g\|_{L^2}<\infty$ ($j=0,\ldots, \ell+1$, see also Proposition 3.4(1) of \cite{Markov2}). Combining this bound on $\N^{-1/2} Y\N^{-1/2}$ with  $\|\N^{1/2}R_z\A\Phi\|\le C\alpha^{-1/2}\|\Phi\|_2$ (see Proposition 3.4(2) of \cite{Markov2}) and $\|R_z\|$, $\|S_\zeta\|\le C/\alpha$ (see Proposition 3.3(1) of \cite{Markov2}), we estimate $|\langle \Phi, \A R_z XS_\zeta^3\Psi\rangle|\le C\alpha^{-3} \|\N^{-1/2}X\N^{-1/2}\|\, \|\Phi\|_2\|\Psi\|_2$ and the same upper bound for the second term on the right side of \eqref{2.61}. The third one is estimated by 
$$|\langle \Phi, R_z [X,\A]S_\zeta^3\Psi\rangle|\le C\alpha^{-3} \|\N^{-1/2}[X,\A]\N^{-1/2}\|\, \|\Phi\|_2\|\Psi\|_2.
$$
For the fourth term on the right side, we take $\ell = 1$ to get $|\langle \Phi
, R_z YR_z  X S_\zeta^3\Psi\rangle|\le C\alpha^{-3} \|\N^{-1/2}X\N^{-1/2}\|\Phi\|_1\|\Psi\|_1$ (we also use the bound $\|\N^{1/2}R_z\N^{1/2}\|\le C$, {\em c.f.} Proposition 3.3(2) of \cite{Markov2}). All together, we obtain
\begin{equation}
\big| \partial_\alpha  \langle \Phi	, R_z  X S_\zeta^3\Psi\rangle\big|\le C\alpha^{-3}  \|\Phi\|_2\|\Psi\|_2. 
\label{2.62}
\end{equation}
We integrate in $\alpha$, using that $|\langle\Phi, R_zXS^3_\zeta\Psi\rangle|_{\alpha=1}\le C\|\Phi\|\, \|\Psi\|$, to obtain
\begin{equation}
	\big| \langle \Phi	, R_z  X S_\zeta^3\Psi\rangle\big|\le C(1+\alpha^{-2})  \|\Phi\|_2\|\Psi\|_2. 
	\label{2.63}
\end{equation}
Now we use the bound \eqref{2.63} to estimate the three first terms on the right side of \eqref{2.61} from above by $C(1+\alpha^{-2})\|\Phi\|_3\|\Psi\|_3$, provided that $\|\N^{-1/2}[[X,\A],\A]\N^{-1/2}\|<\infty$. The norm is now $\|\cdot \|_3$ due to the presence of the operator $\A$ in the first two terms of the right side of \eqref{2.61}. We use $\ell = 2$ to estimate the last term on the right side of \eqref{2.61} and hence we get \eqref{2.62} with $\alpha^{-3}$ replaced by $\alpha^{-2}$. Thus \eqref{2.63} holds with $\alpha^{-2}$ replaced by $\alpha^{-1}$. We now repeat the process one more time, using the latest bounds in \eqref{2.61} and integrating to obtain $| \langle \Phi	, R_z  X S_\zeta^3\Psi\rangle |\le C \|\Phi\|_3\|\Psi\|_3$. It is assumed that $\ell=3$ and $\|\N^{-1/2}[[[X,\A],\A],\A]\N^{-1/2}\|<\infty$.  

This concludes the proof of Lemma \ref{lemma2.1} and hence that of Theorem \ref{theorem1}.\qed

\subsection{Proof of Proposition \ref{prop1.1}}
\label{sec2.4}

We first show Lemma \ref{lemma1.1} and \ref{lemma1.0} detailing some representation independent bounds on polynomials of creation, annihilation and Weyl operators. Then we give the proof of Proposition \ref{prop1.1}, starting at \eqref{2.74} below.

The number operator on the Fock space ${\mathcal F}$ \eqref{F} is given by
\begin{equation}
	\Nh=d\Gamma(\bbbone_{\mathcal F}) =\int_{{\mathbb R}^3} a^*(k) a(k)d^3k.
	\label{c33}
\end{equation}
We often write $\Nh$ for $\bbbone_{{\mathbb C}^N}\otimes \Nh$.  {\em In order to alleviate the notation, in this section we will use the symbol $N$ for the number operator $\widehat N$, even though $N$ also denotes the dimension of the system Hilbert space. }

\begin{lem}
	\label{lemma1.1}
	For any $K\in {\mathcal X} \cup{\mathcal P}$ and any $\epsilon>0$, we have $\|Ke^{-\epsilon N}\|<\infty$.
\end{lem}

{\em Proof of Lemma \ref{lemma1.1}.\ } Let $K=a^{(\sigma_1)}(f_1)\cdots a^{(\sigma_\ell)} (f_\ell)\in{\mathcal P}$, where $\sigma_j\in\{\pm 1\}$ and $a^{(+1)}(f)=a^*(f)$, $a^{(-1)}(f)=a(f)$. Let $\xi$ be a function defined on the integers and define $\xi(N)=\sum_{n\ge 0} \xi(n) P_{N=n}$, where $P_{N=n}$ is the spectral projection of $N$.  The domain of $\xi(N)$ consists of vectors $\psi$ for which the series $\sum_{n\ge 0} \xi(n) P_{N=n}\psi$  converges strongly. Since $P_{N=n}a^*(f)=a^*(f)P_{N=n-1}$ we have $\xi(N)a^*(f)= a^*(f)\xi(N+1)$  and similarly $\xi(N) a(f) = a(f)\xi(N-1)$, so in short, $\xi(N) a^{(\sigma)}(f) = a^{(\sigma)}(f) \xi(N+\sigma)$. Taking $\xi(N)=(N+\ell)^{1/2}$,
\begin{eqnarray}
	K &=& a^{(\sigma_1)}(f_1)(N+\ell)^{-1/2}\cdot  (N+\ell)^{1/2}a^{(\sigma_2)}(f_2)\cdots a^{(\sigma_\ell)}(f_\ell)\nonumber\\
	&=&
	a^{(\sigma_1)}(f_1)(N+\ell)^{-1/2}\cdot  a^{(\sigma_2)}(f_2)\cdots a^{(\sigma_\ell)}(f_\ell)\big(N+\ell+\textstyle\sum_{j=2}^\ell\sigma_j\big)^{1/2}.
	\label{c42}
\end{eqnarray}
We used (and will use below) the symbols $\cdots$ and $\cdot$ for the operation of multiplication of operators. 
Using $(N+\ell)^{-1/2}(N+\ell)^{1/2}=\bbbone$ next to every $a^{(\sigma_j)}(f_j)$ and pulling $(N+\ell)^{1/2}$ to the right as in \eqref{c42} gives 
\begin{equation}
	K =  a^{(\sigma_1)}(f_1)(N+\ell)^{-1/2}\cdots a^{(\sigma_\ell)}(f_\ell)(N+\ell)^{-1/2}\cdot B,
	\label{c34}
\end{equation}
where 
\begin{eqnarray}
	B &=& (N+\ell)^{1/2}(N+\ell+\sigma_{\ell-1})^{1/2} (N+\ell+\sigma_{\ell-1}+\sigma_{\ell-2})^{1/2}\cdots (N+\ell+\textstyle\sum_{j=2}^\ell\sigma_j)^{1/2}\nonumber\\
	&\le& (N+2\ell)^{\ell/2}.
	\label{35}
\end{eqnarray}
We know that $\|a^{(\sigma)}(f)(N+1)^{-1/2}\|\le \|f\|_{L^2}$, see {\em e.g.} \cite{BR,MLnotes}. It follows from \eqref{c34} and \eqref{35},  
\begin{equation}
	\|Ke^{-\epsilon N}\| \le \|f_1\|_{L^2}\cdots\|f_\ell\|_{L^2}\  \sup_{n\in{\mathbb N}}(n+2\ell)^{\ell/2}e^{-\epsilon n}.
	\label{c43}
\end{equation}
Now for $0<\epsilon <1/4$ (which we can take without loss of generality), we have
\begin{equation}
	\sup_{n\in{\mathbb N}}(n+2\ell)^{\ell/2}e^{-\epsilon n} \le \Big(\frac{\ell}{2\epsilon}\Big)^{\ell/2}
	\label{c44}
\end{equation}
and so it follows from \eqref{c43} and \eqref{c44} that the statement of Lemma \ref{lemma1.1} holds for $K$ a product of creation and annihilation operators and hence for all $K\in{\mathcal P}$. Next let 
\begin{equation}
K= e^{\i \sum_{r=1}^R B_r \otimes a^\sharp(f_r) } \in{\mathcal X}.
\label{X}
\end{equation} 
We expand the exponential,
\begin{equation}
K e^{-\epsilon N}= \sum_{\ell=0}^\infty\frac{\i^\ell}{\ell!} \big[ \sum_{r=1}^R B_r\otimes a^\sharp(f_r) \big]^\ell e^{-\epsilon N}.
	\label{c41}
\end{equation}
We then multiply out the product $[\sum_{r=1}^R B_r\otimes a^\sharp(f_r)]^\ell$ into a sum of $R^\ell$ terms, each term being an element  of the form $H\otimes Q\in{\mathcal P}$, where $H\in{\mathcal B}({\mathbb C}^N)$ and $Q$ is a product of $\ell$ creation and annihilation operators. By \eqref{c43}, \eqref{c44}, $\|Qe^{-\epsilon N}\|\le C^\ell\,  (\ell/(2\epsilon))^{\ell/2}$ for some constant $C$. The norm of the general term in the series \eqref{c41} is bounded above by
\begin{equation}
	\Big\| \frac{\i^\ell}{\ell!} \big[\sum_{r=1}^R B_r\otimes a^\sharp(f_r)\big]^\ell e^{-\epsilon N}\Big\| \le C^\ell \frac{1}{\ell !}\Big(\frac{\ell}{2\epsilon}\Big)^{\ell/2}
\end{equation}
for another constant $C$ (depending on $R$) and the series converges for any $\epsilon>0$ (and any $R$), as is easily established using {\em e.g.} the ratio test. This shows that \eqref{c41} is a bounded operator. This concludes the proof of Lemma \ref{lemma1.1}. \qed

The domain of definition of the operator $e^{\epsilon N}$, denoted  ${\rm Dom}(e^{\epsilon N})$,  is a dense set in ${\mathcal F}$.

\begin{lem}
	\label{lemma1.0}
	Given any $\epsilon>0$, all operators in $\mathfrak C$ are well defined on ${\rm Dom}(e^{\epsilon N})$.
\end{lem}

{\em Proof of Lemma \ref{lemma1.0}.\ } To control products of operators, we note that $\forall s\in{\mathbb R}$,  
\begin{eqnarray}
	e^{s N} a^\sharp (f) e^{-s N} &=&  e^{\pm s}a^\sharp(f)\label{c34.1}\\
	e^{s N} e^{\i \sum_{r=1}^R B_r\otimes a^\sharp(f_r)} e^{-s N} &=& e^{\i \sum_{r=1}^R e^{\pm s}\,  B_r\otimes a^\sharp(f_r)},
	\label{34}
\end{eqnarray}
where the $+s$ exponent  is present for $a^\sharp =a^*$ and the $-s$ exponent for $a^\sharp =a$. Equations \eqref{c34.1}, \eqref{34} show that 
\begin{equation}
	e^{s N} {\mathfrak C} e^{-s N} = {\mathfrak C},\qquad \forall s\in{\mathbb R}.
	\label{c38}
\end{equation}
Let now $K_\alpha\in{\mathcal X} \cup{\mathcal P}$, $\alpha=1,\ldots \nu$. Then $\forall \epsilon>0$,
\begin{eqnarray}
	\lefteqn{K_1\cdots K_\nu}\nonumber\\
	&=& K_1 e^{-\epsilon N}\, (e^{\epsilon N}K_2 e^{-2\epsilon N}) \, (e^{2\epsilon N}K_3e^{-3\epsilon N}) \cdots (e^{(\nu-1)\epsilon N}K_\nu e^{-\nu\epsilon N})e^{\nu\epsilon N}.
	\label{c39}
\end{eqnarray}
For $\ell\ge 1$ integer, $e^{(\ell-1) \epsilon N}K_\ell e^{-\ell\epsilon N} = e^{(\ell-1)\epsilon N}K_\ell e^{-(\ell-1)\epsilon N}e^{-\epsilon N}\equiv K_\ell(\epsilon) e^{-\epsilon N}$, where $K(\epsilon)\in{\mathcal X} \cup{\mathcal P}$ is the $K$ with all single-particle functions $f,g$ replaced by $e^{(\ell-1)\epsilon}f$ and $e^{-(\ell-1)\epsilon}g$ for creation and annihilation operators, respectively, according to \eqref{c34.1}, \eqref{34}. As shown in Lemma \ref{lemma1.1}, we have $\|K_\ell(\epsilon)e^{-\epsilon N}\|<\infty$ for any $\epsilon>0$. It follows that 
\begin{equation}
	K_1\cdots K_\nu =K_1 e^{-\epsilon N}K_2(\epsilon)e^{-\epsilon N}K_3(\epsilon)e^{-\epsilon N}\cdots K_\nu(\epsilon) e^{-\epsilon N}\, e^{\nu\epsilon N}
\end{equation}
is a well defined operator on ${\rm Dom}(e^{\nu\epsilon N})$, for any $\epsilon >0$. This completes the proof of Lemma \ref{lemma1.0}.
\qed

{\bf Proof of Proposition \ref{prop1.1}.\ }
To prove Proposition \ref{prop1.1}, we consider
\begin{equation}
\rho_{\s,\beta}\otimes\omega_{\r,\beta}(K^*e^{\i t H_\lambda} A e^{-\i t H_\lambda} K) = \langle \Omega_{\s\r,\beta,0},  \pi(K^*) e^{\i t L_\lambda} \pi(A) e^{-\i t L_\lambda} \pi(K)\Omega_{\s\r,\beta,0}\rangle
\label{2.74}
\end{equation}
for some $K\in{\mathfrak C}$ and $A\in {\mathcal O}$. We only need to show the following result.

\begin{lem}
\label{lemma4.4}
We have  $\|\pi(K)\Omega_{\s\r,\beta,0}\|<\infty$ for any $K\in\mathfrak C$. 
\end{lem}

{\em Proof of Lemma \ref{lemma4.4}.}
According to \eqref{c45} the operator $\pi(a^*(f))$ is the sum of a creation plus an annihilation operator on the Fock space ${\mathcal H}_\r$, \eqref{n48}, and so we have $\|\pi_\beta(a^*(f))(\N+1)^{-1/2}\|<\infty$. Here, $\N$ is the number operator on the Fock space ${\mathcal H}_\r$, \eqref{n48},
\begin{equation}
	\N = d\Gamma(\bbbone_{{\mathcal H}_\r}).
	\label{N}
\end{equation}
We can then proceed entirely analogously to the proof of Lemma \ref{lemma1.1} to show that for all $K\in{\mathcal P}$, we have $\|\pi(K)e^{-\epsilon\N}\|<\infty$, for any $\epsilon>0$.  Next, let $K$ be given by \eqref{X}. From \eqref{2.17}, \eqref{n44.2}, \eqref{c45}  we get 
\begin{equation}
	\pi\big(B_r\otimes a^\sharp(f_r) \big) = B_r \otimes\bbbone_{\s}\otimes \big( a_\beta^*(f_{r,+}) +a_\beta(f_{r,-})  \big),
\end{equation}
for functions $f_{r,+}$, $f_{r,-} \in L^2({\mathbb R}\times S^2)$. It follows that 
\begin{equation}
	\pi\big(e^{\i \sum_{r=1}^R B_r\otimes a^\sharp(f_r)}\big) = e^{\i \sum_{r=1}^R B_r\otimes\bbbone_\s\otimes [ a_\beta^*(f_{r,+}) +a_\beta(f_{r,-})]}. 
	\label{c53}
\end{equation}
Expanding the exponential as in \eqref{c41} and proceeding exactly as in the proof of Lemma \ref{lemma1.1} we see that $\|\pi(e^{\i \sum_{r=1}^R B_r\otimes a^\sharp(f_r)})e^{-\epsilon \N}\|<\infty$ for all $\epsilon>0$.  

So far in this proof, we have shown that 
\begin{equation}
\|\pi(K)e^{-\epsilon \N}\|<\infty,\qquad \forall \epsilon>0,\quad \  \forall K\in{\mathcal X}\cup{\mathcal P}.
\label{2.78}
\end{equation} 
To see that $\|\pi(K_1\cdots K_\nu)\Omega_{\s\r,\beta,0}\|<\infty$ we proceed as in \eqref{c39}, inserting factors $\bbbone = e^{-\ell \epsilon \N} e^{\ell \epsilon \N}$, 
\begin{equation}
\pi(K_1\cdots K_\nu)
	= \pi(K_1) e^{-\epsilon \N}\, (e^{\epsilon \N} \pi(K_2) e^{-2\epsilon \N})  \cdots (e^{(\nu-1)\epsilon \N}\pi(K_\nu) e^{-\nu\epsilon \N})e^{\nu\epsilon \N}.
	\label{c39-1}
\end{equation}
All that is left to do is showing that for any $\ell\ge 1$ integer,
\begin{equation} 
\|e^{(\ell-1)\epsilon\N}\pi(K)e^{-\ell\epsilon \N}\|<\infty,\qquad \forall \epsilon>0, \ \forall K\in {\mathcal X}\cup{\mathcal P}.
\label{2.79}
\end{equation}
We do this as above in Lemma \ref{lemma1.0}. Indeed, we have for all $s\in\mathbb R$,
\begin{eqnarray}
e^{s\N} a_\beta^\sharp(v) e^{-s\N} &=& e^{\pm s} a_\beta(v),\qquad v\in L^2({\mathbb R}\times S^2)\nonumber\\
e^{s\N} \pi\big(e^{\i \sum_{r=1}^R B_r\otimes a^\sharp(f_r)}\big) e^{-s\N} &=& e^{\i \sum_{r=1}^R B_r\otimes\bbbone_\s\otimes [ e^s a_\beta^*(f_{r,+}) +e^{-s} a_\beta(f_{r,-})]},
	\label{c53-1}
\end{eqnarray}
where again, $+s$ and $-s$ are for the creation and the annihilation operator, respectively, as in \eqref{c34.1}, \eqref{34}. The bound \eqref{2.79} now follows from \eqref{2.78} just as in the proof of Lemma \ref{lemma1.0}. This shows that 
\begin{equation}
\|\pi(K)e^{-\epsilon \N}\|<\infty,\quad \|e^{(\ell-1)\epsilon \N}\pi(K)e^{-\ell\epsilon \N}\|<\infty,\qquad \forall \ell\in{\mathbb N}, \epsilon >0, K\in{\mathfrak C}.
\label{2.80}
\end{equation}
This concludes the proof of Lemma \ref{lemma4.4} and hence the proof of Proposition \eqref{prop1.1} is complete.\qed

\subsection{Set of regular vectors $\mathcal D$ and proof of \eqref{c16.1}}
\label{estKsect}

As in \cite{Markov2} we define the norms $\|\cdot\|_j$,  for $j=0,1,2,\ldots$, on $\mathcal H_{\rm GNS}$ \eqref{3.2}: 
\begin{equation}
\|\phi\|_j= \|(1+\bar \A^2)^{j/2}\phi\|, \qquad \A=d\Gamma(\i\partial_u),\qquad  \bar \A = P^\perp_\r \A P^\perp_\r|_{{\rm Ran}P_\r^\perp}.
\label{norm}
\end{equation} 
We have $\|\phi\|_k\le\|\phi\|_\ell$ for $0\le k\le \ell$. The dense set ${\mathcal D}\subset{\mathcal H}_{\rm GNS}$ is given by 
\begin{equation}
{\mathcal D} = \{\phi\in{\mathcal H}_{\rm GNS}\ :\ \|\phi\|_3<\infty \}.
\label{n67.1}
\end{equation}
We now verify that \eqref{c16.1} holds, where the two vectors $\phi$, $\psi$ are given by 
\begin{eqnarray}
\phi&=&\pi(K)\pi'(e^{\beta H_0/2} K e^{-\beta H_0/2})\Omega_{\s\r,\beta,0},\nonumber\\
\psi &=& \pi(A)\Omega_{\s\r,\beta,0}.
\label{c15+}
\end{eqnarray}
We first show that $\|\phi\|_3<\infty$. We have
\begin{eqnarray}
e^{\beta H_0/2} (B\otimes a^\sharp(f) )e^{-\beta H_0/2} &=& e^{\beta H_\s/2} Be^{-\beta H_\s/2} \otimes a^\sharp(e^{\pm\beta |k|/2} f) \label{c47.1} \\
e^{\beta H_0/2} e^{\i \sum_{r=1}^R B_r\otimes a^\sharp(f_r)} e^{-\beta H_0/2} &=& e^{\i \sum_{r=1}^R e^{\beta H_\s/2}B_re^{-\beta H_\s/2}\otimes a^\sharp(e^{\pm\beta |k|/2} f_r)}
\label{c47}
\end{eqnarray}
where the plus sign in the  exponentials in \eqref{c47.1}, \eqref{c47} is present if $a^\sharp = a^*$ and the minus sign if $a^\sharp =a$. Note that $e^{\pm\beta H_\s/2}$ is bounded and $\e^{\pm\beta|k|/2}f_r\in L^2$, so that the operators \eqref{c47.1} and \eqref{c47} have the same structure as the operators in ${\mathfrak C}$ (except that here, the single-particle functions $f$ appear with the weights $e^{\pm \beta|k|/2}$). It follows that analyzing $\phi$ in \eqref{c15+} reduces to analyzing vectors of the form $\phi'=\pi(K)\pi'(K')\Omega_{\s\r,\beta,0}$, where $K, K'\in{\mathfrak C}_0$, and where $\mathfrak C_0$ is the algebra defined by ({\em c.f.} \eqref{c35-1})
\begin{equation}
{\mathfrak C}_0 ={\rm LinSpan} \big\{ K_1\cdots K_n\ :\ K_j\in{\mathcal X}_0 \cup {\mathcal P}_0 \ :\  n\in{\mathbb N}\big\},
\label{c49}
\end{equation}
where ${\mathcal X}_0$ and ${\mathcal P}_0$ are as $\mathcal X$ and $\mathcal P$ given in \eqref{c36} and \eqref{c37}, but with $L^2_{\rm cor}$ replaced by $L^2_{\rm obs}$ (we do not assume $e^{\beta |k|}f\in L^2({\mathbb R}^3, d^3k)$). Using that $J\N=\N J$ and $\N\Omega_{\s\r,\beta,0}=0$, we get
\begin{eqnarray}
\pi(K)\pi'(K') \Omega_{\s\r,\beta,0}&=& \pi(K)J\pi(K')J \Omega_{\s\r,\beta,0}\nonumber\\
&=& \pi(K)e^{-\epsilon \N} Je^{\epsilon \N}\pi(K')e^{-2\epsilon\N}J \Omega_{\s\r,\beta,0}.
\label{c48}
\end{eqnarray}
The relation \eqref{c48} together with \eqref{2.80} (which holds for $K\in{\mathfrak C}_0$) then shows immediately that $\|\pi(K)\pi'(K')\Omega_{\s\r,\beta,0}\|<\infty$, for all $K,K'\in{
\mathfrak C}_0$. In other words, $\|\phi\|_0<\infty$, where this is the norm \eqref{norm} with $j=0$ of the vector $\phi$,  \eqref{c15+}. We now show that $\|\phi\|_3<\infty$. Writing simply $\A$ for $\bar \A$ (see \eqref{norm}) in the following argument, we have
\begin{equation}
\|\phi\|^2_j = \langle\phi, (1+\A^2)^j\phi\rangle\le C\langle \phi, (1+\A^{2j})\phi\rangle = C\big( \|\phi\|^2 +\|\A^j\phi\|^2\big),
\label{c50}
\end{equation}
for a constant $C$ independent of $\phi$. It follows that $\|\phi\|_j\le C ( \|\phi\|+\|A^j\phi\|)$. We thus only need to show that $\|\A^3\phi\|<\infty$, that is, we need to show that
\begin{equation}
\|\A^3\pi(K)\pi'(K')\Omega_{\s\r,\beta,0}\|<\infty,\quad \mbox{ for all $K,K'\in{\mathfrak C}_0$. }
\label{c51}
\end{equation}
We have for all $\alpha\in\mathbb R$ (see also \eqref{c53}),
\begin{equation}
e^{\i\alpha \A}  \pi(e^{\i\sum_{r=1}^R B_r\otimes a^\sharp(f_r)})e^{-\i\alpha \A} = e^{\i \sum_{r=1}^R B_r\otimes\bbbone_\s \otimes[  a_\beta^*(e^{-\alpha\partial_u}f_{r,+})+ a_\beta(e^{-\alpha\partial_u}f_{r,-})]}.
\label{c54}
\end{equation}
Applying the operation $-\i\partial_\alpha|_{\alpha=0}$ to the left side of \eqref{c54} gives the commutator of $\A$ with $ \pi(e^{\i\sum_{r=1}^R B_r\otimes a^\sharp(f_r)})$. To calculate the corresponding right hand side, we recall that well-known formula for an operator family $Y(\alpha)$ depeding on $\alpha\in\mathbb R$,
\begin{equation}
\partial_\alpha e^{Y(\alpha)} = \int_0^1 e^{sY(\alpha)} \big(\partial_\alpha Y(\alpha)\big) e^{(1-s)Y(\alpha)}ds. 
\label{c52}
\end{equation}
Thus $-\i\partial_\alpha|_{\alpha=0}$ applied to both sides of \eqref{c54} results in the expression
\begin{equation}
\A \pi(e^{\i\sum_{r=1}^R B_r\otimes a^\sharp(f_r)}) -  \pi(e^{\i\sum_{r=1}^R B_r\otimes a^\sharp(f_r)})\A = {\mathcal C}
\label{c55},
\end{equation}
where the commutator is
\begin{eqnarray}
\lefteqn{ {\mathcal C}=-\int_0^1 \pi(e^{\i s \sum_{r=1}^R B_r\otimes a^\sharp(f_r)})}\nonumber\\
&&\quad \times  \Big(\sum_{r=1}^R B_r\otimes\bbbone_\s\otimes [  a^*_\beta(\partial_u f_{r,+}) +a_\beta(\partial_uf_{r,-})] \Big)
\pi(e^{\i (1-s) \sum_{r=1}^R B_r\otimes a^\sharp(f_r)})ds.\ \
\label{c58}
\end{eqnarray}
Of course,  $\mathcal C$ is not a bounded operator, as usual, but by \eqref{2.80}, we have 
\begin{equation} 
\|{\mathcal C}e^{-\epsilon\N}\|<\infty,\quad \|e^{\epsilon \N} {\mathcal C} e^{-2\epsilon \N}\|<\infty,\qquad \forall \epsilon >0.
\label{c56}
\end{equation} 
Next, using that $\A a^*_\beta(v)-a^*_\beta(v) \A = a^*_\beta(\i\partial_u v)$ for any $v\in L^2({\mathbb R}\times S^2)$, we see that for any polynomial $P\in{\mathcal P}_0$, we have $\A \pi(P)-\pi(P)\A = {\mathcal C}$ for an operator $\mathcal C$ satisfying again \eqref{c56}. Let now $L_1,\ldots, L_n$ each be an operator belonging to ${\mathcal X}_0\cup{\mathcal P}_0$. Then
$ \A \pi(L_1\cdots L_n) = \pi(L_1) \A\pi(L_2\cdots L_n) + {\mathcal C_1}\pi(L_2\cdots L_n)$, and keeping up commuting $\A$ to the right through the operators $L$ gives
\begin{equation}
\A\pi( L_1\cdots L_n) = \pi(L_1\cdots L_n) \A + {\mathcal C},
\end{equation}
where ${\mathcal C}$ satisfies \eqref{c56}. Now since $\pi'(\cdot) = J\pi(\cdot)J$ and $J\A=\A J$, we get with the same argument that 
\begin{equation}
\A\pi^\sharp( L_1)\cdots \pi^\sharp(L_n) = \pi^\sharp(L_1)\cdots \pi^\sharp(L_n) \A + {\mathcal C},
\label{c57}
\end{equation}
with ${\mathcal C}$ satisfying \eqref{c56}. Here, each $\pi^\sharp$ individually is either $\pi$ or $\pi'$. Then by linearity, the formula \eqref{c57} is also correct if each $L_j\in{\mathfrak C}_0$ (as opposed to being just an element of ${\mathcal X}_0\cup{\mathcal P}_0$). 

We now combine \eqref{c57} with the fact that $\A\Omega_{\s\r,\beta,0}=0$ to conclude that \eqref{c51} holds true for $\A^3$ replaced by $\A$. In other words, we have $\|\phi\|_1<\infty$. But applying $\A$ to both sides of \eqref{c57} gives 
\begin{eqnarray}
\A^2 \pi^\sharp( L_1)\cdots \pi^\sharp(L_n) &=& \A\pi^\sharp(L_1)\cdots \pi^\sharp(L_n) \A + \A{\mathcal C}_1\nonumber\\
&=& \pi^\sharp(L_1)\cdots \pi^\sharp(L_n) \A^2+{\mathcal C}_2\A + \A{\mathcal C}_1.
\label{c59}
\end{eqnarray}
We can commute $\A$ through ${\mathcal C}_1$, $\A{\mathcal C}_1={\mathcal C}_1\A+{\mathcal C}'$. The operator ${\mathcal C}'$ once again will satisfy $\|{\mathcal C}'e^{-\epsilon \N}\|<\infty$, any $\epsilon>0$. This is easily seen since ${\mathcal C}_1$ is a sum of products of elements in ${\mathcal P}_0$ and operators of the form \eqref{c58}. One may then proceed as above to show the bound $\|{\mathcal C}'e^{-\epsilon \N}\|<\infty$, any $\epsilon>0$. This shows that $\|\phi\|_2<\infty$. We repeat the procedure and apply $A$ to \eqref{c59} to conclude that $\|\phi\|_3<\infty$. The only limitation on the number of times we can repeat the procedure is that successive derivatives $\partial_u$ of all functions $f_{r,+}$, $f_{r,-}$ involved should stay in $L^2({\mathbb R}\times S^2)$. 

Each additional application of $A$ requires one more such derivative to be $L^2$. For $\|\phi\|_j<\infty$ we need the derivatives up to and including order $j$ to be square integrable. This condition for $j=0,\ldots,3$ is guaranteed by taking functions $f(k)$ from the set $L^2_{\rm cor}$, see the point (a) at the beginning of Section \ref{sec:oa}.

This shows that $\|\phi\|_3<\infty$. Next we need to show that $\|\bar L_\lambda\phi\|_3<\infty$ (see \eqref{c16.1}). We have
\begin{equation}
\A^3 L_\lambda = L_0\A^3 +3\i \N \A^2 +\lambda\big( I\A^3+ 3I_1\A^2 +3 I_2\A +I_3\big),
\end{equation}
where $I_k$ is the $k$ fold commutator of $I$ with $\A$. We have $\| I(\N+1)^{-1/2}\|<\infty$ and $\| I_k(\N+1)^{-1/2}\|<\infty$. It follows that $\|\bar L_\lambda \phi\|_3<\infty$ provided that 
\begin{equation}
\|\N \A^j\phi\|<\infty, \ j=0,\ldots,3 \quad \mbox{and} \quad \|L_\r \A^3\phi\|<\infty.
\label{2.99}
\end{equation}
By the above arguments, we know that $\|e^{\epsilon\N}\phi\|<\infty$
 and $\|e^{\epsilon \N} \A^3\phi\|<\infty$ for all $\epsilon>0$, so the bounds involving $\N$ in \eqref{2.99} hold. To show that $\A^3\phi\in{\rm Dom}(L_\r)$ we repeat the argument after \eqref{c55} above. Instead of commuting $\A$ through elements of ${\mathfrak C}_0$, now we have commute $L_\r$ through them. 
 We have $L_\r a^\sharp_\beta(v)-a^\sharp_\beta(v)L_\r= \pm a_\beta^\sharp(uv)$, where the plus sign is for $a_\beta^\sharp=a_\beta^*$ and $uv$ is the function $u\cdot v(u,\Sigma)$.  The multiplication by $u$ of the functions $v(u,\Sigma)$ preserves the square integrability as guaranteed by the the definition of $L^2_{\rm obs}$, $L^2_{\rm cor}$ (point (a) at the beginning of Section \ref{sec:oa}).  The same argument as above then shows that $\|L_\r \A^3\phi\|<\infty$.
 
We have shown so far that $\phi, \bar L_\lambda\phi\in \mathcal D$. To finish the proof of \eqref{c16.1} we also need to show that $\psi, \bar L_\lambda\psi\in\mathcal D$, where $\psi=\pi(A)\Omega_{\s\r,\beta,0}$, see \eqref{c15}. According to \eqref{3.9} we have 
\begin{equation}
\pi_\beta (W(f))=e^{\i \varphi_\beta(\tau_\beta f)} = e^{\i [a_\beta^*(\tau_\beta f) + a_\beta(\tau_\beta f)]/\sqrt2} \in {\mathfrak C}_0.
\label{2.101}
\end{equation}
It follows that ${\mathcal O}\subset{\mathfrak C}_0$ and hence showing $\psi,\bar L_\lambda\psi\in\mathcal D$ is a special case of the proof that $\phi,\bar L_\lambda\phi\in\mathcal D$. This completes the proof of \eqref{c16.1}. \qed

\subsection{Level shift operators}
\label{FGRsect}

The level shift operators $\Lambda_e$ are defined in \eqref{n98-1}. As $I$ contains two terms, see \eqref{n40}, $\Lambda_e$ is the sum of four terms,
\begin{eqnarray}
	\lefteqn{
\Lambda_e = }\label{n99}\\
 &&  -P_e \{ G\otimes \bbbone_\s\otimes \varphi_\beta(\tau_\beta g) \} (L_0-e+\i 0_+)^{-1} \{G\otimes \bbbone_\s\otimes \varphi_\beta(\tau_\beta g)\}P_e\nonumber\\
&& +P_e \{ G\otimes \bbbone_\s\otimes \varphi_\beta(\tau_\beta g) \} (L_0-e+\i 0_+)^{-1} \{\bbbone_\s\otimes {\mathcal C}G{\mathcal C}\otimes \varphi_\beta(e^{-\beta u/2}\tau_\beta g)\}P_e\nonumber\\
&&+P_e \{\bbbone_\s\otimes {\mathcal C}G{\mathcal C}\otimes \varphi_\beta(e^{-\beta u/2}\tau_\beta g)\} (L_0-e+\i 0_+)^{-1} \{ G\otimes \bbbone_\s\otimes \varphi_\beta(\tau_\beta g) \}P_e\nonumber\\
&&-P_e \{\bbbone_\s\otimes {\mathcal C}G{\mathcal C}\otimes \varphi_\beta(e^{-\beta u/2}\tau_\beta g)\} (L_0-e+\i 0_+)^{-1} \{\bbbone_\s\otimes {\mathcal C}G{\mathcal C}\otimes \varphi_\beta(e^{-\beta u/2}\tau_\beta g)\}P_e.
\nonumber
\end{eqnarray}
The partial trace over the reservoir part is calculated using the formula
\begin{eqnarray}
\lefteqn{
P_\r \varphi_\beta(F)  (L_0-e+\i 0_+)^{-1}  \varphi_\beta(G) P_\r =\tfrac12 P_\r a_\beta(F)  (L_0-e+\i 0_+)^{-1}  a^*_\beta(G) P_\r}\nonumber\\
&& \qquad  \qquad \qquad = \tfrac12 P_\r \int_{{\mathbb R}\times S^2} \bar F(u,\Sigma) G(u,\Sigma) (L_\s-e+u+\i 0_+)^{-1}du d\Sigma,
\label{n100}
\end{eqnarray}
valid for any two functions $F,G\in L^2({\mathbb R}\times S^2)$. 
One obtains from \eqref{n99}
\begin{eqnarray}
\lefteqn{
\Lambda_e =}\label{n101}\\
&& -\tfrac12 P_{\s,e} (G\otimes\bbbone_\s) \int_{{\mathbb R}\times S^2} \frac{\big|g(|u|,\Sigma)\big|^2}{|1-e^{-\beta u}|} (L_\s-e+u+i 0_+)^{-1}  u^2 du d\Sigma\  (G\otimes\bbbone_\s) P_{\s,e}\nonumber\\
&&+\tfrac12 P_{\s,e} (G\otimes\bbbone_\s) \int_{{\mathbb R}\times S^2} \frac{e^{-\beta u/2}\big|g(|u|,\Sigma)\big|^2}{|1-e^{-\beta u}|} (L_\s-e+u+i 0_+)^{-1}  u^2 du d\Sigma\  (\bbbone_\s\otimes{\mathcal C}G{\mathcal C}) P_{\s,e}\nonumber\\
&&+\tfrac12 P_{\s,e} (\bbbone_\s\otimes {\mathcal C}G{\mathcal C}) \int_{{\mathbb R}\times S^2} \frac{e^{-\beta u/2}\big|g(|u|,\Sigma)\big|^2}{|1-e^{-\beta u}|} (L_\s-e+u+i 0_+)^{-1}  u^2 du d\Sigma\ (G\otimes\bbbone_\s)  P_{\s,e}\nonumber\\
&&-\tfrac12 P_{\s,e} (\bbbone_\s\otimes {\mathcal C}G{\mathcal C}) \int_{{\mathbb R}\times S^2} \frac{\big|g(|u|,\Sigma)\big|^2}{|e^{\beta u}-1|} (L_\s-e+u+i 0_+)^{-1}  u^2 du d\Sigma\ (\bbbone_\s\otimes {\mathcal C}G{\mathcal C})  P_{\s,e}.\nonumber
\end{eqnarray}
The condition that $\Lambda_e$ should have simple spectrum can then be verified for concrete cases, where $g, G$ are given explicitly. Note that if $e$ is a simple eigenvalue of $L_\s$, then $P_{\s,e}$ has rank one and so $\Lambda_e$ has automatically simple spectrum. Note also that $e=0$ is always a degenerate eigenvalue of $L_\s$. The explicit form of the level shift operators for several models can be found in \cite{MBS, Markov2}. Additional information on their structure is provided in \cite{Mlso}.

\bigskip
{\bf Acknowledgements.\ } The author thanks two anonymous referees for examining this work and giving helpful comments. The author was supported by a {\em Discovery Grant} from NSERC, the {\em National Sciences and Engineering Research Council of Canada}.


\end{document}